\RequirePackage{fix-cm}
\documentclass[twocolumn]{svjour3}          
\smartqed  
\usepackage{graphicx}
\usepackage[utf8]{inputenc}
\usepackage{caption}
\usepackage{subcaption}
\usepackage{epsfig}
\usepackage{amsmath}
\usepackage[version=4]{mhchem}
\usepackage{amsmath}
\usepackage{array}
\usepackage{pict2e}
\usepackage[percent]{overpic}
\usepackage{siunitx}
\usepackage{color}
\usepackage{longtable,tabularx}
\usepackage{graphicx}
\usepackage{placeins}
\usepackage{tikz}
\usepackage{multicol}

\newcommand{\at}[2][]{#1|_{#2}}
\usepackage{amsfonts}

\usepackage{color}
\usepackage[normalem]{ulem}

\begin{document}

\title{Lift Coefficient Estimation for a Rapidly Pitching Airfoil}
\subtitle{}


\author{Xuanhong An         \and
        David R. Williams \and
        Jeff D. Eldredge  \and
        Tim Colonius
}

\institute{Xuanhong An \at
              MAE Department, Princeton University, Princeton, NJ 08544, USA  \\
              \email{xuanhong@princeton.edu}
           \and
           David R. Williams \at
              MMAE Department, Illinois Institute of Technology, Chicago, IL, 60616, USA
            \and  
            Jeff D. Eldredge \at
            MAE Department, University of California Los Angeles, Los Angeles, CA, 90095, USA
            \and
            Tim Colonius \at
            MCE Department, California Institute of Technology, Pasadena, CA, 91125, USA
}

\date{Received: date / Accepted: date}

\maketitle

\begin{abstract}
We develop a method for estimating the instantaneous lift coefficient on a rapidly pitching airfoil \textcolor{black}{that uses a small number of  pressure sensors and a measurement of the angle of attack}. The approach assimilates four surface pressure measurements with a modified nonlinear state space model (Goman-Khrabrov model) through a Kalman filter. The error of lift coefficient estimates based only on a weighted-sum of the measured pressures are found to be noisy and biased, which leads to inaccurate estimates. The estimate is improved by including the predictive model in an conventional Kalman filter. The Goman-Khrabrov model is shown to be a linear parameter-varying system and can therefore be used in the Kalman filter without the need for linearization. Additional improvement is realized by modifying the algorithm to provide more accurate estimate of the lift coefficient. The improved Kalman filtering approach results in a bias-free lift coefficient estimate that is more precise than either the pressure-based estimate or the Goman-Khrabrov model on their own.  The new method will enable performance enhancements in aerodynamic systems whose performance relies on lift.
\end{abstract}

\section{Introduction}
\label{intro}


Obtaining accurate, real-time estimates of the instantaneous lift acting on airfoils will enable performance enhancements in many practical applications, particularly in unsteady flow environments. Active control systems for aircraft, rotorcraft, and wind turbines could benefit from direct knowledge of the aerodynamic loads rather than relying upon identification of deviations of the vehicle from the desired trajectory. If the instantaneous lift and lift-history are known, then controllers can be designed that will alleviate gusting flow effects, reduce the unsteady loading, and improve flight control in turbulent environments.  

Examples of practical applications that could benefit from real-time lift estimation include helicopter rotor blades, wind turbine blades, aircraft wings and their control surfaces. Bio-inspired flyers rely on pitching maneuvers for enhanced lift at low Reynolds number flight conditions \cite{articleB} \cite{article}. In each case the airfoils experience time-varying angles of attack that can lead to stall or dynamic stall that can reduce system performance. 

Potential performance enhancements for these applications include increased flight speed for rotorcraft, enhanced gust alleviation capabilities for aircraft, improved flight trajectory tracking, and improved maneuverability.  For example, during forward flight helicopters must continuously adjust the inclination angle of the rotor blades during the rotation cycle to maintain balanced lift on the port and starboard sides of the helicopter.  The ability to control the asymmetry of lift plays a major role in limiting the helicopter forward flying speed. Similarly, adjustments to flight vehicle attitude during landing approaches or when flying through gusts require rapid control surface deflections, even though the aircraft themselves may not change pitch attitude. 

Generally speaking, performance can be enhanced if the instantaneous lift coefficient is continually estimated even at post-stall angle of attack. One approach to estimating the lift coefficient is to measure the angle of attack and use an aerodynamic model to predict the lift coefficient.  A second method is to use surface pressure sensors with a spatially-weighted averaging scheme to measure the normal force coefficient.  The normal force coefficient can be corrected to the lift coefficient by using the angle of attack.  Some advantages and disadvantages to these approaches are described next.


\subsection{Lift estimation using low-dimensional models}

High-fidelity methods, such as large-eddy simulation are too computationally intensive for use in real-time controllers. Meanwhile, reduced order approaches such as the classical Wagner model \cite{wagner1925entstehung} and its multi-plate extension \cite{theodorsen1979general} employ potential flow theory and the Kutta condition, and thus cannot account for flow separation that resulting unsteadiness associated with vortex shedding.

The ability of a low-order model to capture the transition between the attached and separated flow is important. Hemati, et al.\cite{hemati2016parameter}, Brunton, et al. \cite{brunton2014state}, Dawson \cite{dawson2015data} and Provost, et al. \cite{le2018sindy} introduced different linear parameter-varying (LPV) models that showed good performance for aerodynamic load tracking on rapidly pitching wings.  However, these models are purely data-driven and the models' physical insight remains to be investigated. Goman \& Khrabrov \cite{goman1994state} (G-K model) proposed a state-space model, utilizing the nonlinear static lift measurement as a forcing term. This model is capable of predicting the lift force during rapid maneuvers, and the model's stability is guaranteed by constraining the values of the time constant. Grimaud \cite{grimaud2014energy} and Williams, et al. \cite{williams2015dynamic} modified the G-K model in such a way that the lift is a function of a single inner state variable and angle of attack. 
The evolution of the G-K type models is described in a review article by Williams and King \cite{williams2018alleviating}   

Despite its good performance, the G-K model is a first-order model that neglects some other important higher-order features in the flowfield, such as, trailing-edge vortex separation, natural vortex shedding and some other unmodeled disturbances.  Peters, et al. \cite{peters1995finite} introduced a second-order model that provided higher fidelity lift estimates at the expense of increased computation time. 

\textcolor{black}{It is important to recognize that the low-dimensional models require angle of attack as input, which is additional information about the state of the airfoil that is not necessarily required by the pressure measurement approach.  The pressure measurement approach is discussed next.}

\subsection{Lift estimation with surface pressure}
A different approach to estimating the lift coefficient is to use real-time surface pressure measurements.  Ideally, if a sufficient number of pressure sensors are available, then the sum of spatially weighted surface pressure measurements will give a good estimate of the instantaneous vector force (without the contribution of skin friction) from which the lift can be obtained. 
A recent example is a wind turbine application by Bartholomay, et al. \cite{wes-2020-91}, who investigated the ability of a small number of surface pressure sensors on a wind turbine blade to identify the instantaneous lift.  The measured lift was then used in a feed-forward controller to reduce the unsteady loads acting on the turbine blade.

In practical applications there are limitations on the number of pressure sensors that can be installed on the surface. The accuracy of the lift estimate will be reduced as the number of sensors is reduced, but depending on the application some success has been demonstrated.  An, et al.\cite{an2017response} showed that with a limited number of pressure sensors, it is possible to project the state variable (pressure distribution along the entire airfoil) onto its sub-space (sparse pressure measurements), which then leads to colored noise for the lift that can, in turn, be estimated by the sparse pressure measurements. Such colored noise can even be nonlinear. We refer to this colored noise as the biased error for the remainder of this paper, in contrast to the white measurement noise. The biased error can be problematic when the Kalman filter is implemented. Unlike white noise, with little knowledge of the (time-varying) biased error, it is difficult for a Kalman filter to reduce the biased error. Therefore, a new way of coupling the model and the measurement is proposed in the present work that enables the Kalman filter to reduce the biased error.

{Some investigators \cite{dawson2015data} \cite{le2018sindy} \cite{darakananda2018data} employed the Kalman filter \cite{kalman1960new} to assimilate pressure measurements into low-order models (other than the G-K model) for estimation of the real-time aerodynamic loads variation in response to different types of wing maneuvers.} It has been shown that the white noise coming from the pressure measurements can be reduced by the Kalman filter, but the colored noise (biased error) is yet to be investigated.  


In the following sections, we begin by using four surface pressure sensors to estimate the lift coefficient using a weighted average.  The error in the lift coefficient is shown to have a bias.  To improve the estimate, the data is assimilated with a modified G-K model by using a Kalman filter. We will show that the modified G-K model is in fact an LPV model with a static nonlinear forcing term, so that the model is amenable to conventional Kalman filtering without any linearization of the G-K model. Finally, the combined model is shown to eliminate the bias and produce a lift coefficient estimate that is more precise than either of the G-K model prediction or the weighted-pressure estimate. 


The paper is organized as follows. The experimental setup is described in Sec. \ref{sec:exp}. The pressure-based estimate of the lift coefficient and the associated biased error is discussed in Sec. \ref{sec:pressure_CL}. Next, the derivation of the LPV form of the modified G-K model is given in Sec. \ref{sec:LPV}.  The design of the Kalman filter is discussed in Sec. \ref{sec:LQE}, and the validation of the Kalman filter is given in Sec. \ref{sec:validate}. Finally, the main results are summarized in Sec. \ref{Sec:conc}.     


        
\section{Experimental Setup}\label{sec:exp}
\begin{figure}[ht]
	\begin{subfigure}{0.5\textwidth}
	\includegraphics[width=3.3in]{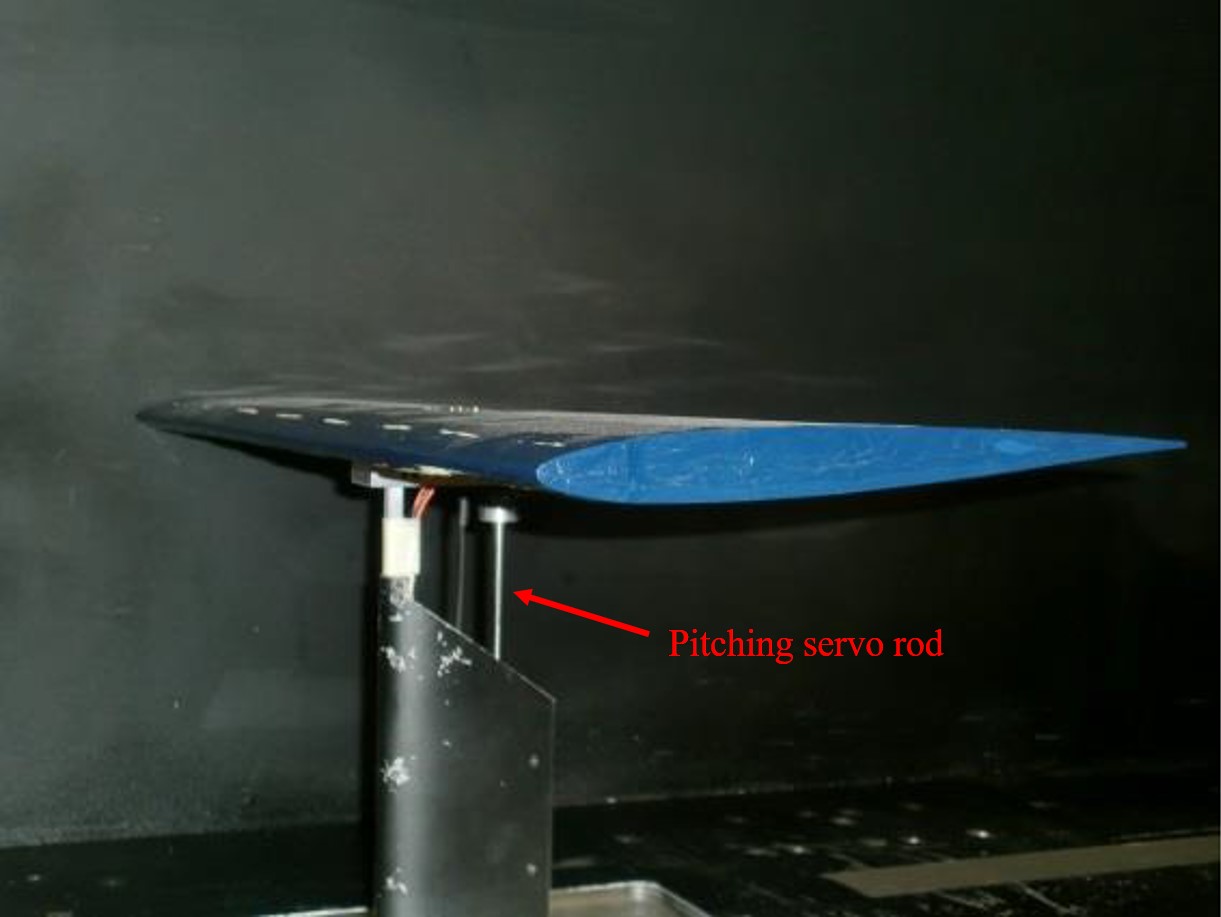} 
		          \caption{The NACA-0009 wing mounted on the pitching mechanism.}
		          \label{fig:wing_tunnel}
	\end{subfigure}
\\
	\begin{subfigure}{0.5\textwidth}
	\centering
	\includegraphics[width=3.3in]{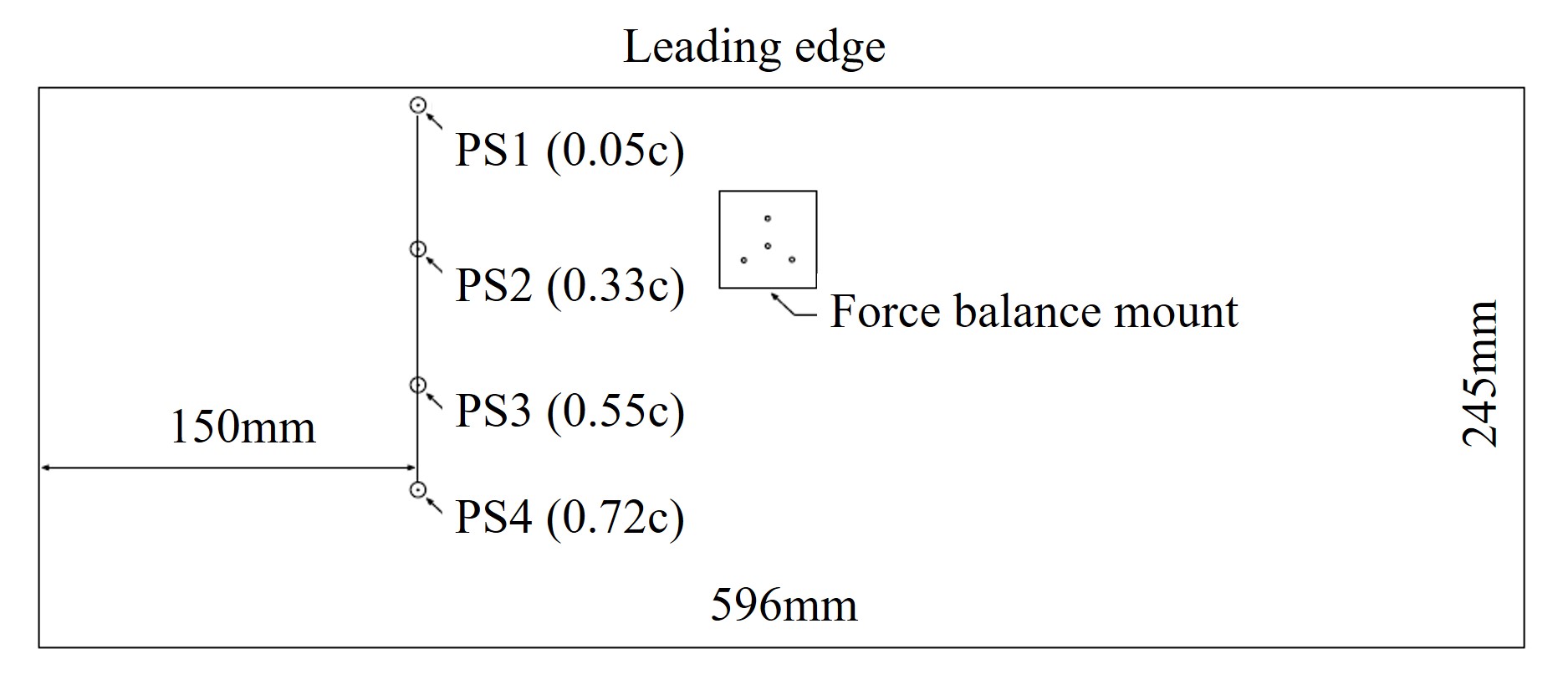} 
		          \caption{The top view of the wing showing pressure sensor and force balance locations.}
		          \label{fig:wing_top}
	\end{subfigure}
	
    \caption{Photo (a) and schematic (b) of the NACA-0009 wing in the test section.}
    \label{fig:exp_setup}		
\end{figure}

The experiments were conducted in the Andrew Fejer Unsteady Flow Wind Tunnel at Illinois Institute of Technology. The test section of the wind tunnel has cross-section dimensions 600 mm $\times$ 600 mm. A nominally two-dimensional NACA-0009 wing with a wingspan $b$ = 596 mm and chord length $c$ = 245 mm was used as the test article (Fig. \ref{fig:wing_tunnel}). The gaps to prevent contact between each wing tip and the sidewalls of the wind tunnel were 2 mm, and the sidewall boundary layer was approximately 20 mm thick.

The freestream speed was $U_{\infty}=3 \ \textrm{m}/\textrm{s}$, corresponding to a convective time, $t_{c} = \frac{c}{U_{\infty}}\approx0.08 \ \textrm{s}$, $t^+=\frac{t}{t_{c}}$, and chord-based Reynolds number 49,000. The freestream turbulence level in the frequency band of 0.1 Hz to 200 Hz was 0.11 percent of the mean flow speed.  The reduced frequency $K$ is defined as $K=\frac{\pi f_{Hz}c}{U_{\infty}}$ where $f_{Hz}$ is the frequency in Hz. 

Direct force measurements acting on the wing were acquired with an ATI Nano-17 force balance.  The force balance was connected to the pitch-plunge mechanism consisting of two computer-controlled Copley servo tubes. The two servo tubes enable the pivot point for the pitching motion to be changed. For the results presented in this paper, the pivot point was at the location, $x_{pivot}/c$ = 0.15. Pitch rates were restricted to 2 Hz or less ($K$ $<$ 0.55) to avoid over-stressing the force balance. Forces were measured with the force balance located inside the model at $30\%$ of the chord, which is the center of gravity of the wing. 

Four pressure (1inchD2P4Vmini) sensors are located on the upper surface of the wing along its chord line. The locations of the force balance and the pressure sensors are shown in Fig. \ref{fig:wing_top}. The lift coefficient is defined as $C_L = \frac{L}{1/2\rho U_{\infty}^2 c}$, where $L$ is the lift and $\rho$ is the air density. In order to subtract the inertial effect of the wing while it is moving, we first measured the lift force as the offset lift ($L_{\textrm{off}}$) on the moving wing at $U_{\infty} = 0$, and then subtract $L_{\textrm{off}}$ from total lift force at $U_{\infty}$. Loosely speaking, the added mass is also part of $L_{\textrm{off}}$. 

\section{Lift coefficient estimation with sparsely distributed pressure sensors}\label{sec:pressure_CL}
Direct measurement of the lift coefficient ($C_L$) is not possible on aircraft during flight, but an estimate of the normal force coefficient can be obtained using surface pressure measurements.  Ideally, if the complete pressure distribution and angle of attack, $\alpha$, are known then the exact lift coefficient can be found.  Practical constraints limit the the number of pressure sensors that can be installed in a wing. To estimate the lift coefficient using a sparse distribution of pressure measurements, it is necessary to use a set of weighting coefficients along with an offset to formulate the following equation \cite{r2014},

\begin{equation} \label{eq:58}
C_{L}(t)=\cos{(\alpha(t))}\sum_{_i=1}^{N+1}w_i p_{i}(t),
\end{equation}
where $C_L(t)$ and $\alpha(t)$ are the instantaneous lift coefficient and angle of attack, respectively, and $p_i(t)$ is the instantaneous pressure at the $i$th of $N$ sensor locations. Furthermore, $w_i$ are weighting coefficients and a constant offset has been added by defining $p_{N+1}(t) \equiv 1$.

In order to calibrate the weights, we collect training data where both $C_L$ and $p_i$ are measured simultaneously, and find the weights that minimize the least-squared error in Eq.~(\ref{eq:58}). Let $P_{i}(t)=\cos{(\alpha(t))}p_{i}(t)$, and collect the equations over a set of discrete times $t_k, \ k=1,2,\ldots, M$ into a matrix form
\begin{equation} \label{eq:59}
\textbf{C}_{\textbf{L}}=\textbf{P} \  \textbf{w},
\end{equation}
where the $(i,k)$th element of the matrix $\textbf{P}$ is $P_i(t_k)$. The best weights are given by the pseudo-inverse
\begin{equation} \label{eq:510}
\textbf{w}=\textbf{P}^T[\textbf{P}\textbf{P}^T]^{-1}\textbf{C}_{\textbf{L}}.
\end{equation}

\begin{figure}[h]
	\centering
    \includegraphics[width=0.5\textwidth]{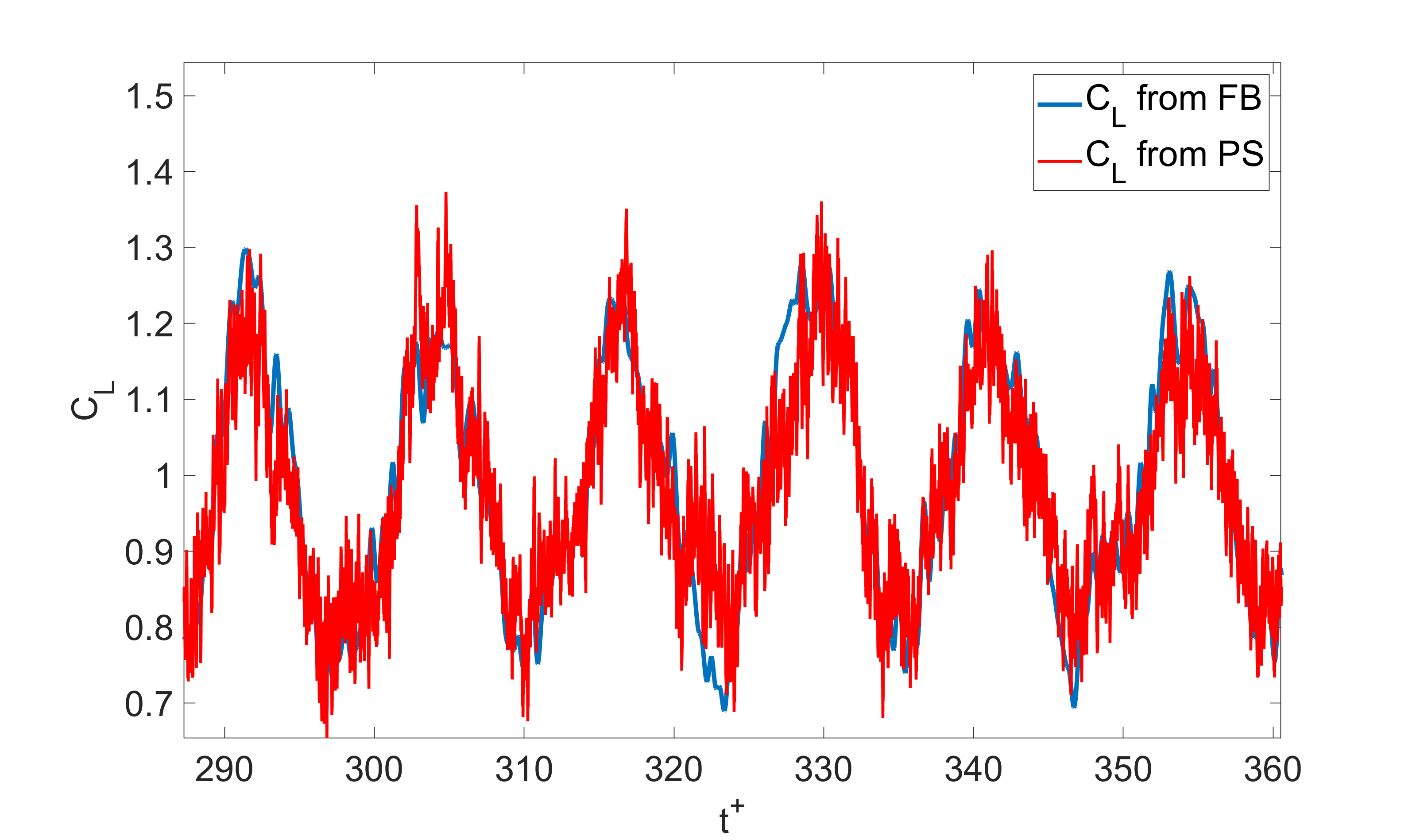}
    \caption{Comparison of $C_L$ measured by the force balance (FB) and $C_L$ estimated by pressure sensors (PS) for the training case. The presetting pitching motion is from $13^o$ to $19^o$ at $K=0.13$} 
    \label{fig:single_CL_CM} 
\end{figure}

In our experiment, there are four pressure sensors on the surface of the airfoil's suction side (Fig. \ref{fig:wing_top}). The reason for placing all four pressure sensors on the suction side is that we focus on positive angles of attack, so that flow structures associated with separation occur on the suction side. A training data set consisting of 4076 data points (time-series data) from an airfoil that pitches from $13^o$ to $19^o$ at $K=0.13$ was used to solve for $\textbf{w}$. The comparison between the $C_L$ estimated by pressure measurements and the 'true' $C_L$ measured by force balance (FB) for the training case is shown in Fig. \ref{fig:single_CL_CM}. The agreement appears reasonable, but the lift inferred from the pressure signals is noisier.



\FloatBarrier

\begin{figure}[h]
	\centering
    \includegraphics[width=0.5\textwidth]{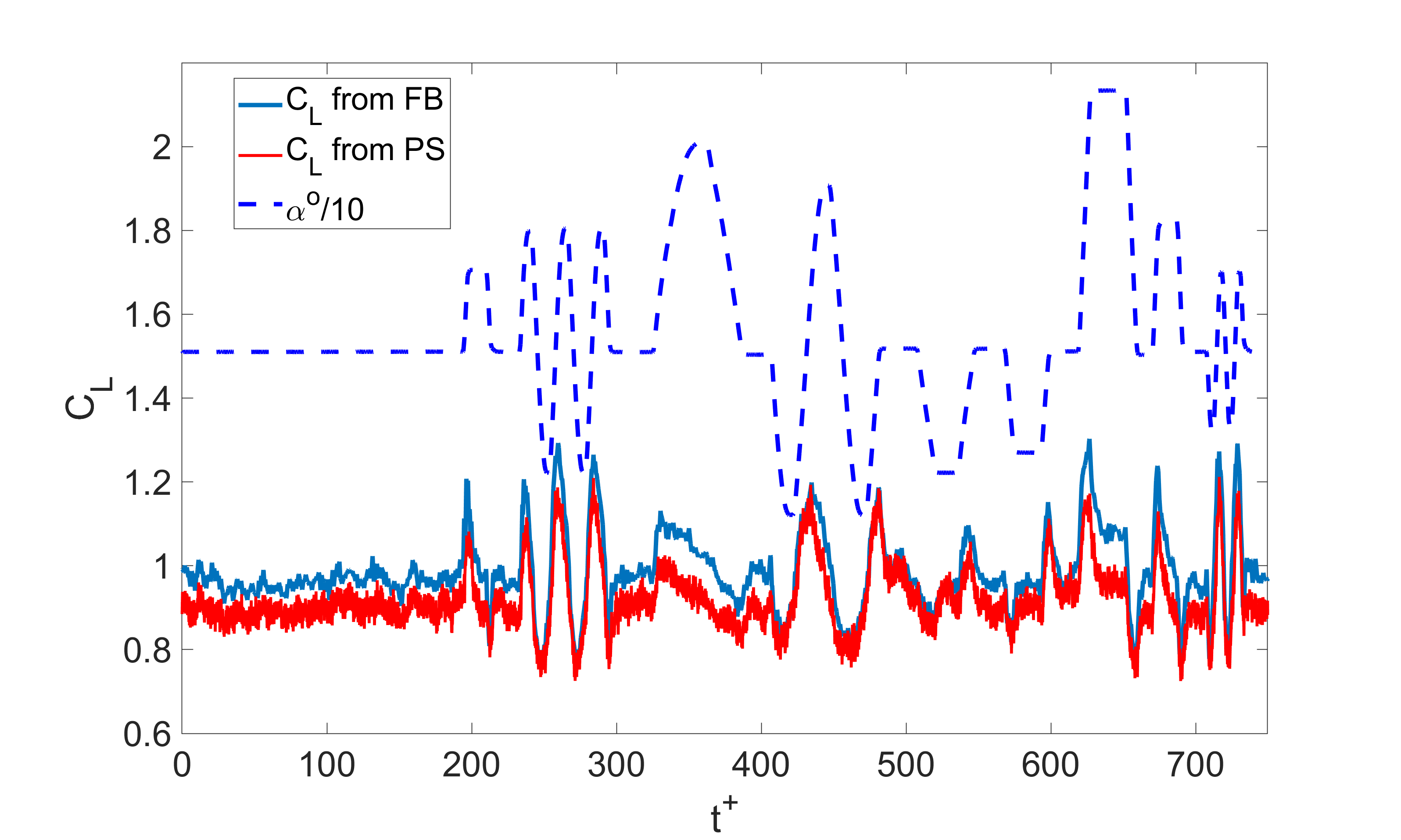}
    \caption{Comparison of $C_L$ measured by the force balance and $C_L$ estimated by pressure sensors for the non-training case.} 
    \label{fig:CL_P_random} 
\end{figure}

The weighted pressure method was then tested on a quasi-random pitching motion. The result is shown in Fig. \ref{fig:CL_P_random}. The quasi-random pitching motion was  constructed by superposing 10 sinusoidal signals with random  initial phases.  The highest reduced frequency of the sinusoidal signals is $K =  0.51$. 
The figure shows that in addition to the noise inherent to the pressure-inferred lift value, there is also an offset (bias) that varies with both the value and rate of change of angle of attack.

\FloatBarrier

\begin{figure}[h]
	\centering
    \includegraphics[width=0.5\textwidth]{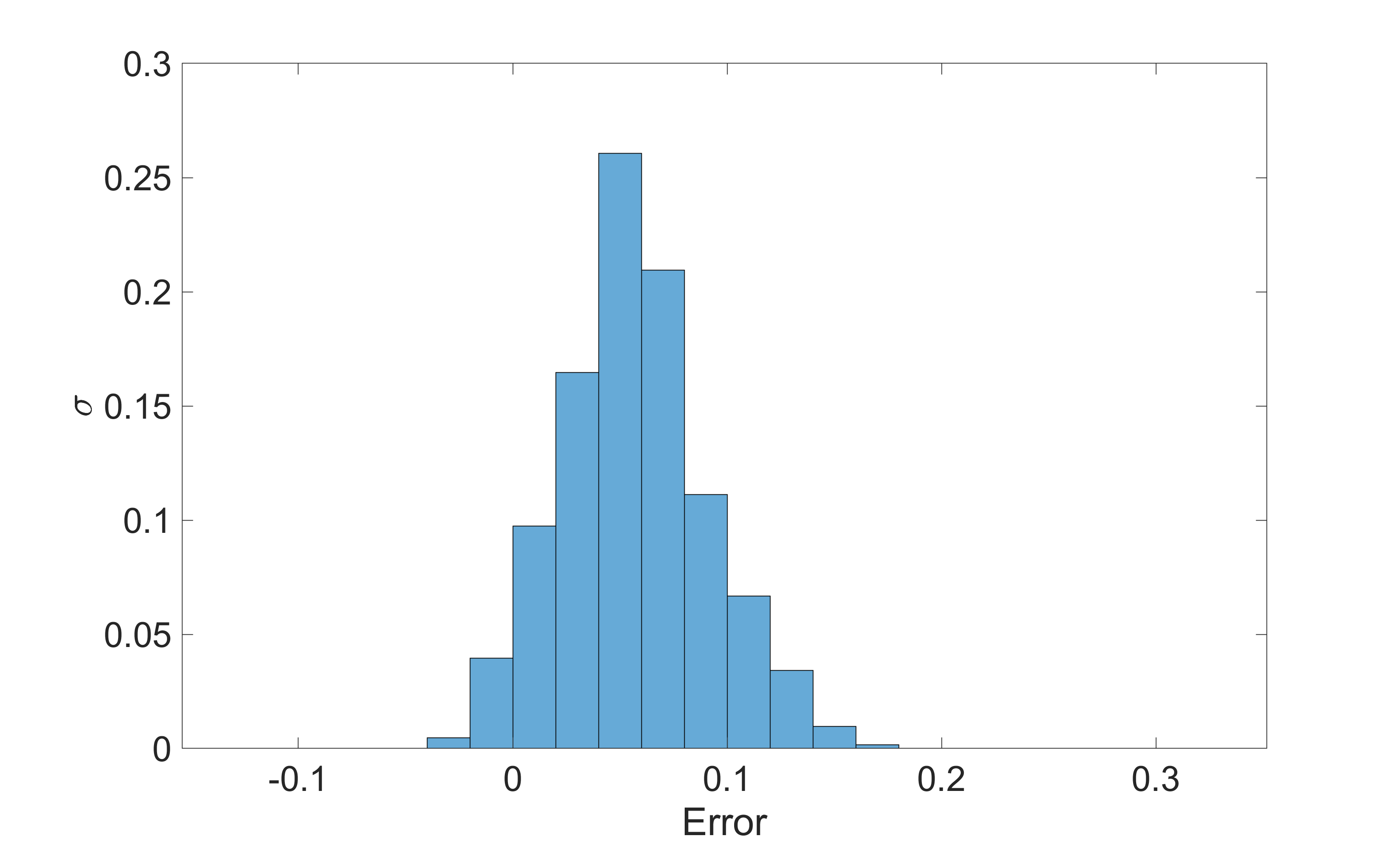}
    \caption{The error between $C_L$ measured by the force balance and $C_L$ estimated by pressure sensors for the random pitching case.} 
    \label{fig:CL_P_err} 
\end{figure}

A histogram of the estimation error is shown in Fig. \ref{fig:CL_P_err}. The error is defined as the (signed) difference between the $C_L$ values measured by the force balance and the pressure measurements. The non-zero mean of the distribution shows that the measurement is biased and the apparent skewness has ramifications for any controller design based on the model. In the histogram, more spread means more white error, and the middle line (bias) of the histograms indicate the biased error (colored noise), that is saying, tall thin strips in the vicinity at Error $=0$ indicates small $C_L$ estimation error. The height, $\sigma$, of each bin is normalized by the total number of counts.

\begin{figure}[h]
	\centering
    \includegraphics[width=0.5\textwidth]{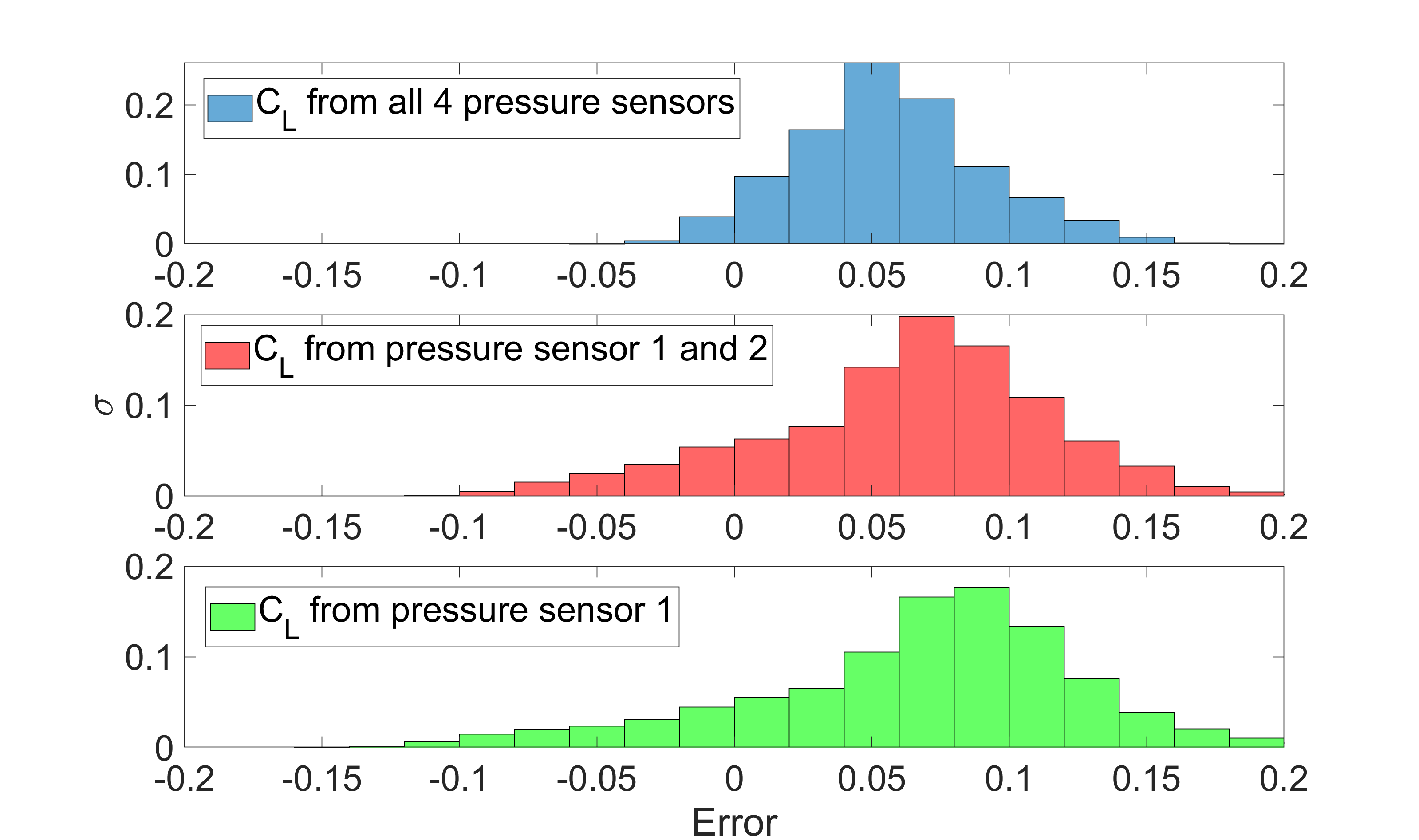}
    \caption{Comparison of the error of $C_L$ measured by the force balance and $C_L$ estimated by different numbers of pressure sensors for the non-training case.} 
    \label{fig:CL_P_err_sensors} 
\end{figure}

In order to further investigate the variance of the estimate, we repeat the analysis using different numbers of pressure sensors in the $C_L$ distribution. The results, shown in Fig. \ref{fig:CL_P_err_sensors}, indicate that the variance increases as the number of sensors is reduced, which shows that there is error cancellation amongst the multiple sensors.  Such error cancellation could be related to both random noise in the individual sensors, or sensing pressure fluctuations that are uncorrelated amongst the sensors (i.e. because their length scale is too small to be simultaneously sensed).  In addition, the offset grows and the distribution is increasingly skewed with fewer sensors.  

These observations are helpful when constructing a quadratic estimator for $C_L$, which we pursue in Sec.~\ref{sec:LQE}.  In the next section, we first address a plant model based on an extension of the G-K approach.

\FloatBarrier
\FloatBarrier
 
\section{A linear parameter-varying model}\label{sec:LPV}

To improve the pressure-based lift estimate we first incorporate a dynamic model for the lift coefficient. The model is a modification of the G-K model discussed above.  In this section, we introduce the model, and show that it corresponds to an LPV system.  After analysing the performance of the model, it will be combined with the pressure-based lift estimate discussed in the previous section and formulated as a linear quadratic estimator in Sec.~\ref{sec:LQE}.

\subsection{The Goman-Khrabrov model}

\begin{figure}[h]
	\centering
    \includegraphics[width=.45\textwidth]{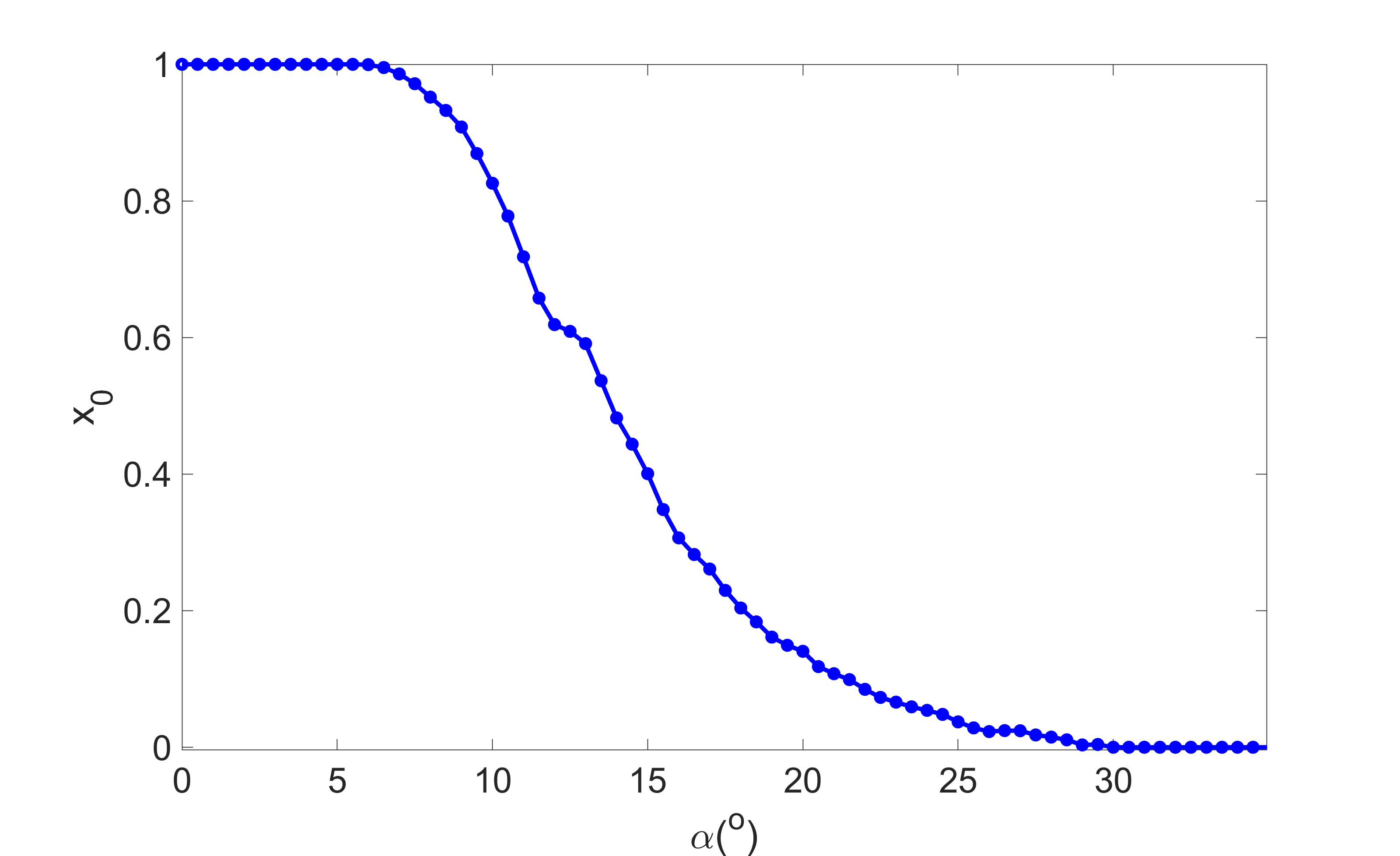}
    \caption{Quasi-steady quantity of the degree of attachment, $x_0(\alpha)$ in the  modified G-K model.} 
    \label{fig:X0_GK} 
\end{figure}

We can divide the unsteady fluid mechanics processes into two groups. The first group refers to the quasisteady effects that vary with the attitude of the objects (angle of attack). The second group is related to the transient aerodynamic effects which are related to delay and relaxation process. The Goman-Khrabrov (G-K) expresses these two groups of effects using a first-order differential equation \cite{goman1994state} . 
The original G-K model is formulated in terms of a dimensionless internal dynamic variable, $x$, that nominally represents the degree of flow attachment over the wing.  Fully attached flow corresponds to $x=1$, and fully separated flow is $x = 0$.  The evolution of $x$ is given by a first-order ODE
\begin{equation} \label{eq:21}
\tau_1 \frac{dx}{dt}+x=x_0(\alpha-\tau_2\dot{\alpha}),
\end{equation}
where $\tau_1$ and $\tau_2$ are empirical time constants.  Observables are then correlated with $x$ and $\alpha$, for example the instantaneous lift coefficient can be expressed as \cite{grimaud2014energy}
\begin{equation} \label{eq:22}
C_L(\alpha,x)=2\pi\alpha(0.4+0.6x)+0.1,
\end{equation}
where the time dependence is inherited through variations in $x$ and $\alpha$. The function $x_0(\alpha)$ represents the degree of separation (measured in $x$ units) during a slow, quasi-steady pitching motion.  The value can be inferred by solving Eq.~(\ref{eq:22}) with an $\alpha$-$C_L$ static map. For the current NACA-0009 airfoil the resulting $x_0(\alpha)$ is shown in Fig. \ref{fig:X0_GK}. Since the computational cost of G-K model is very low, the two time constants are often obtained by running through all the possible values of the time constants to find the values that minimize the mean square error between the model and the training data (eg. a dynamic pitching motion). For the current test conditions, the time lag associated with dynamic stall vortex formation and its convection over the wing is represented by $\tau_2 = 4.375t^+$, and the relaxation time constant is $\tau_1 = 3.75t^+$. The Euler method is used to compute $x(t)$ from Eq.~(\ref{eq:21}) during real-time experiments. 

 

\FloatBarrier

Eq.~(\ref{eq:22}) is determined by trial and error, but can be replaced by a more systematic formulation \cite{williams2016modeling} based on specific static measurements.  In particular, we take
 
\begin{equation} \label{eq:23}
\resizebox{0.45\textwidth}{!}{$
C_L(\alpha,x)=C_1(\alpha(t)-C_3)x(t)+C_2(\alpha(t)-C_4)(1-x(t))
$}
\end{equation}     
where $C_1$ is the $\alpha-C_L$ slope, $\frac{dC_L}{d\alpha}$, when the flow is fully attached, $C_3$ is the zero-lift angle, $C_2$ is the $\alpha-C_L$ slope, $\frac{dC_L}{d\alpha}$ for fully separated flow and $C_2(\alpha-C_4)$ is the $C_L$ value at the smallest $\alpha$ when the flow is fully separated. In what follows we will refer Eqs. \ref{eq:21} and \ref{eq:23} as the modified G-K model (mG-K model).

\subsection{The relation between the mG-K model and an LPV model}

Eq.~(\ref{eq:21}) is a linear, constant-coefficient, ODE for the (scalar) state $x$, while Eq.~(\ref{eq:23}) relates the observable (lift coefficient) to the state through a linear, but non-constant-coefficient (time-varying) expression.  As the coefficients must be determined through real-time measurements (of the angle of attack), the model is thus an LPV system \cite{shamma1992gain}. 


The linearity of Eqs.~\ref{eq:23} in $x$ also allows the mG-K model into a single equation for advancing the lift coefficient, which simplifies the development of the Kalman filter in the next section.  Substituting Eq.~(\ref{eq:23}) into Eq.~(\ref{eq:21}), we obtain
\begin{equation}\label{eq:PVL} 
    {d C_L \over d t} = \left( \frac{\dot g}{g} - \frac{1}{\tau_1} \right) C_L + \left( 
    \frac{g}{\tau_1}  X_0 + \frac{f}{\tau_1} - \frac{\dot g}{g} f  + {\dot f} \right),
\end{equation}
where 
\begin{align}
    X_0 & = x_0(\alpha-\tau_2 \dot{\alpha}) \\
    f(t) & = C_2 \left( \alpha(t) - C_4 \right) \\
    g(t) & = C_1 \left( \alpha(t) - C_3 \right) - f(t).
\end{align}
Upon discretization in time (explicit Euler method), we then obtain
\begin{equation}\label{eq:dynamic0}
    {C_L}(t_{k+1}) = a_k {C_L}(t_k) + b_k
\end{equation}
where
\begin{align}
    a_k &= 1 - \Delta t \ \left( \frac{\dot g (t_k)} {g(t_k)} - \frac{1}{\tau_1} \right), \\
    b_k & = \Delta t \left( 
    \frac{g(t_k)}{\tau_1}  X_0 + \frac{f(t_k)}{\tau_1} - \frac{\dot g (t_k)}{g(t_k)} f(t_k)  + {\dot f (t_k)} \right),
\end{align}
and $\Delta t$ is the time increment.  Note that the coefficients $a_k$ and $b_k$ depend on $\alpha(t_k)$ and $\dot\alpha (t_k)$.  The latter quantity is evaluated as 
\begin{equation}\label{eq:dynamic}
    \dot\alpha (t_k) = \frac{1}{\Delta t} \left( \alpha(t_k) - \alpha(t_{k-1}) \right),
\end{equation}
which is consistent with the $O(\Delta t)$ error invoked in the Euler discretization.

This model can be applied directly to the conventional Kalman filter as a $C_L$ estimator, which we do in Sec.~\ref{sec:LQE}. A more detailed proof of the applicability of Kalman filter on this LPV system is given in the Appendix. In the remainder of this section, we validate the mG-K model for fast ($K \geq 0.05$)  periodic and quasi-random pitching maneuvers.

\subsection{Periodic motion} 

\begin{figure}
	\begin{subfigure}{0.45\textwidth}
	\centering
	\includegraphics[width=1\textwidth]{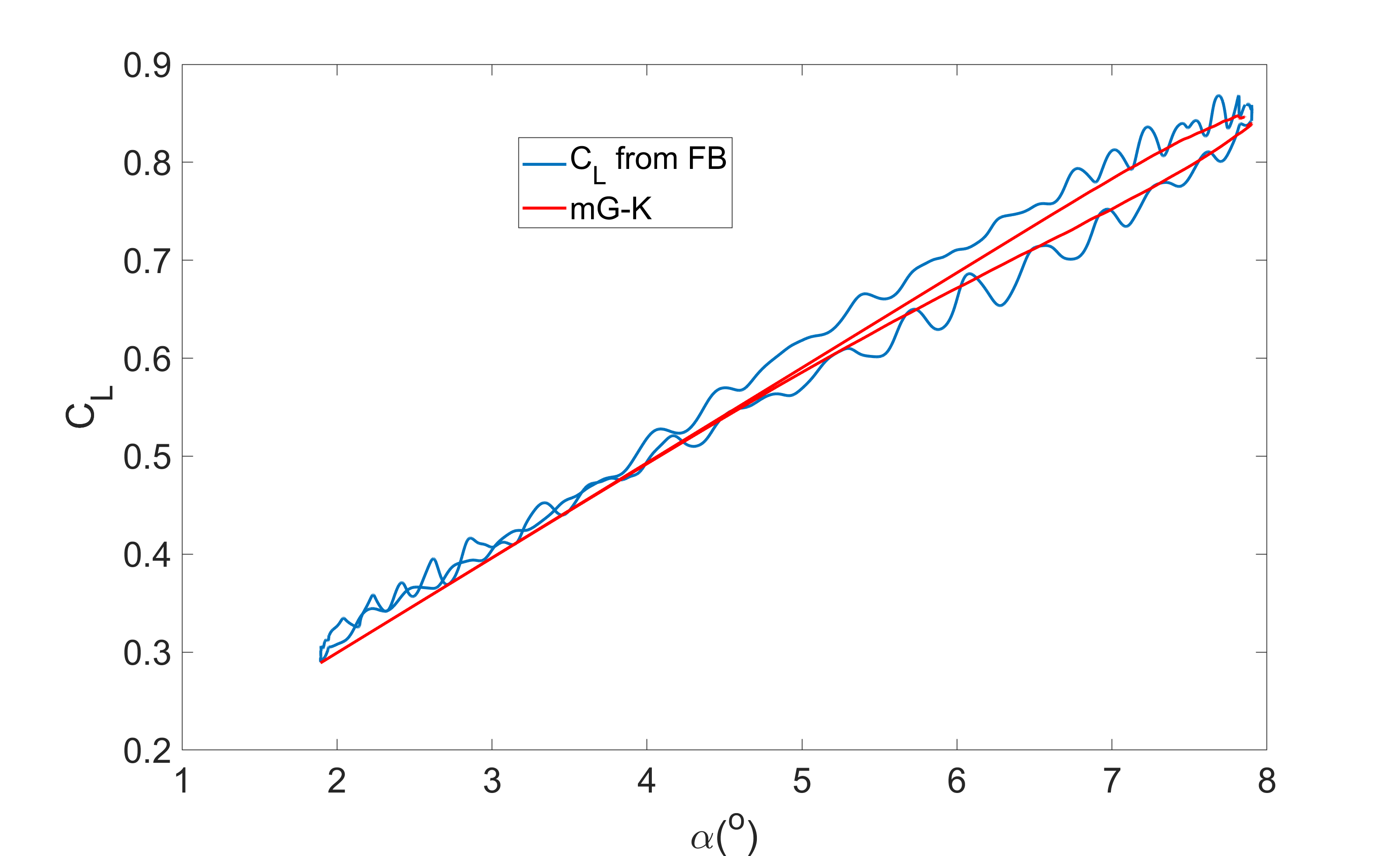} 
		          \caption{$K = 0.05$, $\alpha$ from $2^o$ to $7.8^o$}
		          \label{fig:sin_GK_0.064}
	\end{subfigure}
	
	\begin{subfigure}{0.45\textwidth}
	\centering
	\includegraphics[width=1\textwidth]{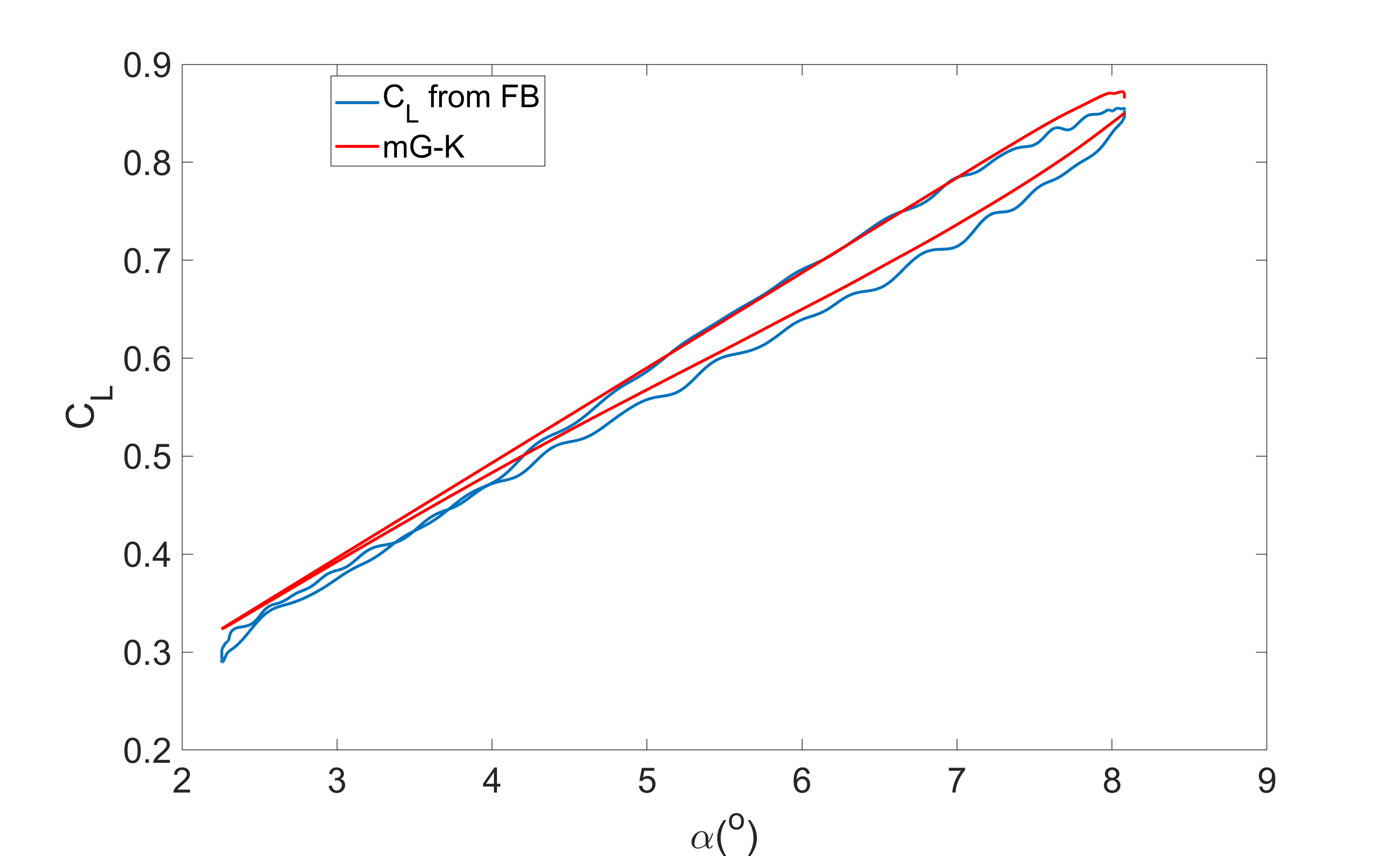} 
		          \caption{$K=0.1$, $\alpha$ from $2.3^o$ to $8^o$}
		          \label{fig:sin_GK_0.1}
	\end{subfigure}
	
	\begin{subfigure}{0.45\textwidth}
	\centering
		\includegraphics[width=1\textwidth]{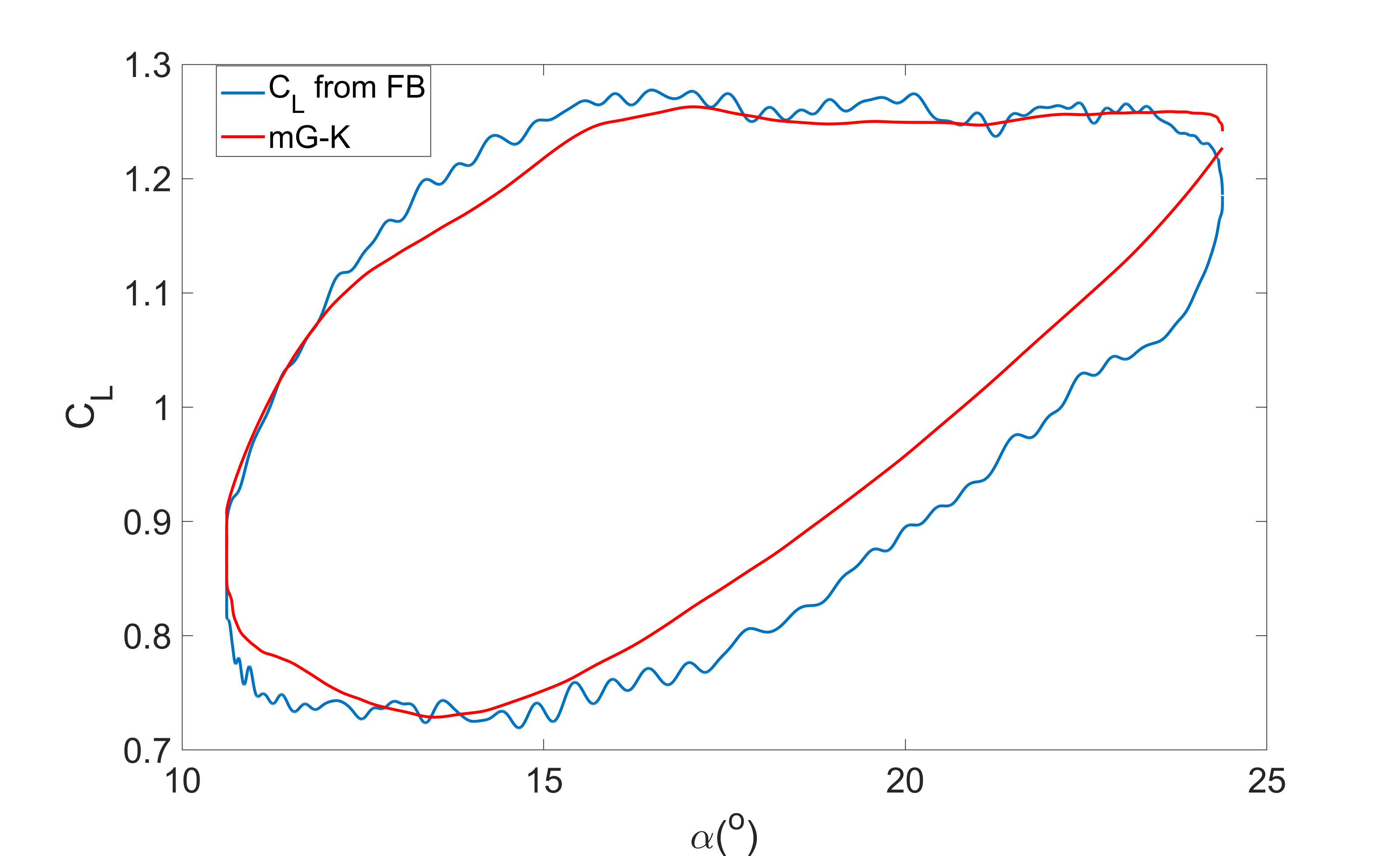} 
			          \caption{$K=0.06$, $\alpha$ from $11^o$ to $24^o$}
			          \label{fig:sin_GK_0.25}
	\end{subfigure}
	
		\begin{subfigure}{0.45\textwidth}
		\centering
			\includegraphics[width=1\textwidth]{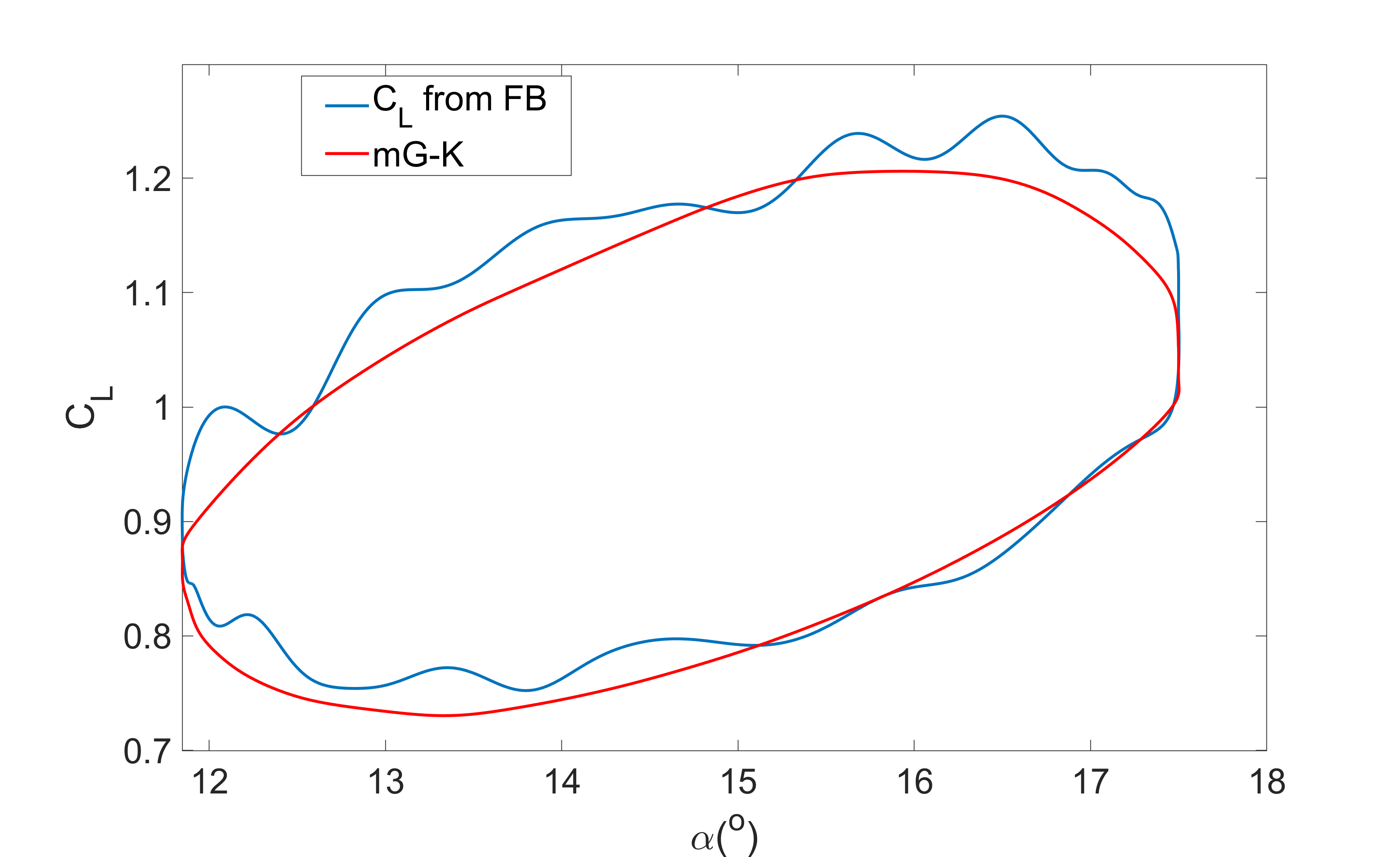} 
				          \caption{$K=0.128$, $\alpha$ from $11.9^o$ to $17.5^o$}
				          \label{fig:sin_GK_0.128}
		\end{subfigure}
		
    \caption{Force balance measured and mG-K modeled $C_L$ for sinusoidal pitching motions.}
    \label{fig:GK_periodic}		
\end{figure}

For a periodic pitching motion the lift coefficient deviates from its quasi-steady values and hysteresis loops are formed. The ability of the mG-K model to predict the lift hysteresis is shown in Fig.~\ref{fig:GK_periodic} for four cases with different pitching frequencies and ranges of $\alpha$ (see figure caption for values).
The mG-K model is capable of tracking the changes in $C_L$ during these periodic pitching motions. Even the dynamic stall in Fig. \ref{fig:sin_GK_0.064} and Fig. \ref{fig:sin_GK_0.1} is captured where the flow is attached in the quasi-steady case shown in Fig. \ref{fig:X0_GK}. 



\FloatBarrier

\subsection{Quasi-random motion} 
\begin{figure}[h]
	\centering
    \includegraphics[width=.5\textwidth]{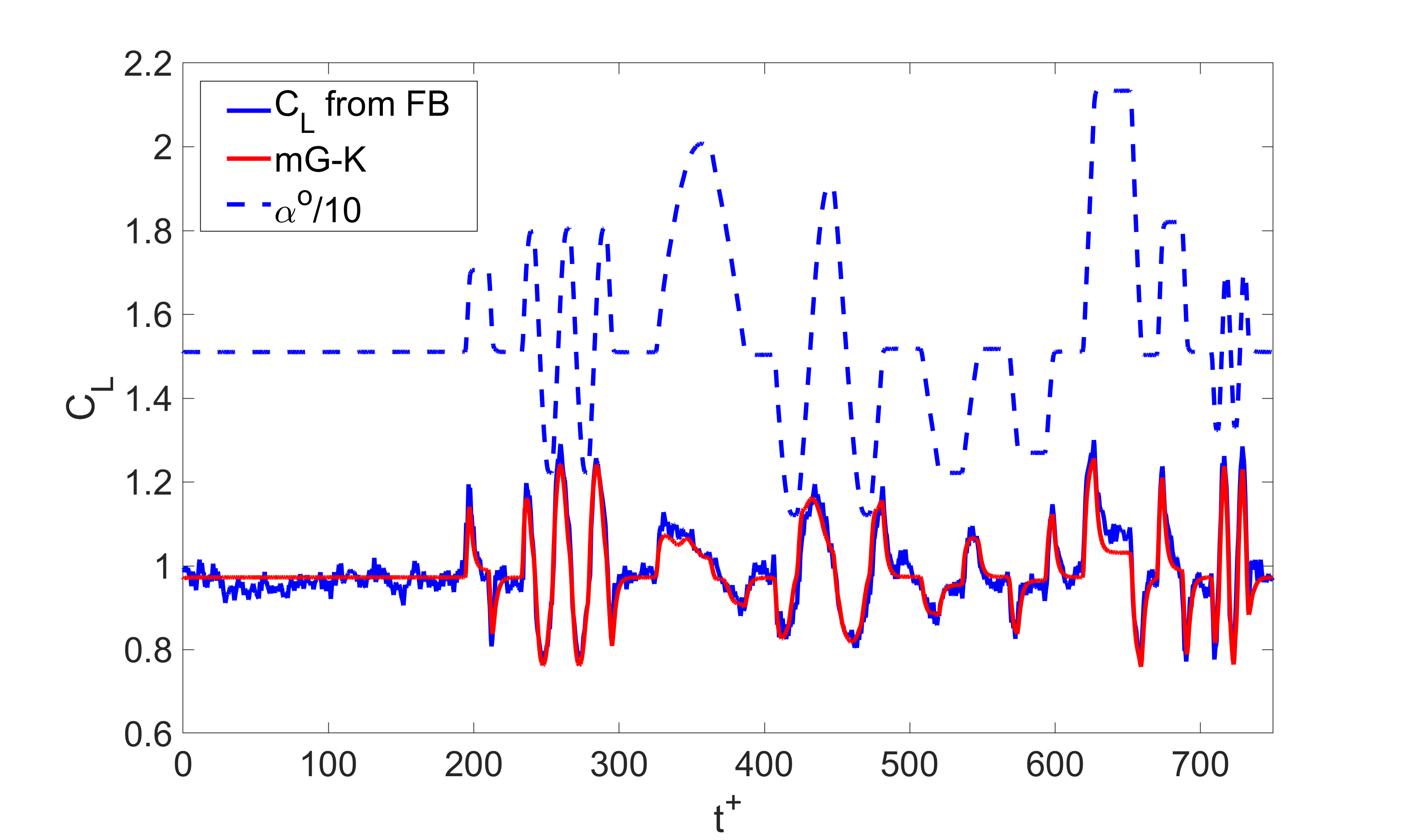}
    \caption{Force balance measured and mG-K modeled $C_L$ for random pitching motions.} 
    \label{fig:GK_random} 
\end{figure}

The ability of the mG-K model to predict the lift coefficient variation produced by a quasi-random pitching motion is demonstrated in Fig. \ref{fig:GK_random}, where the
maneuver is the same as the one used in Sec. \ref{sec:pressure_CL}.  The mG-K model prediction closely tracks the experimental data, and the correlation coefficient between them is 0.956. 

However, some errors still exist due to fluctuations associated with turbulence not captured by the mG-K model. For instance, $C_L$ predicted that the mG-K model remains a constant from $0t^+$ to $200t^+$ since $\alpha$ remains constant, whereas the force balance shows small turbulent fluctuations about the constant value. This result again, suggests that a Kalman filter utilizing both the mG-K model and sparse pressure measurements will be beneficial for accurate $C_L$ estimation. 


\section{Linear Quadratic Estimator (Kalman filter) design}\label{sec:LQE}

\begin{figure}[h]
	\centering
    \includegraphics[width=0.45\textwidth]{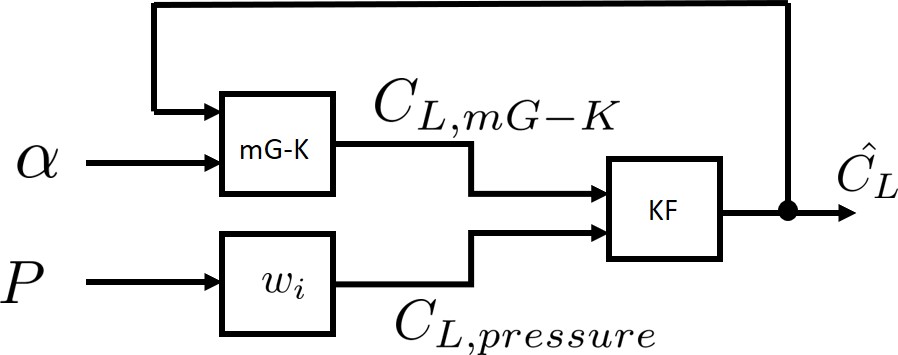}
    \caption{Schematic of the conventional Kalman filter design using Eq.~(\ref{eq:dynamic}) and the pressure sensors ($C_{L,pressure}$ from Eq.~(\ref{eq:58})) measurements separately.}
    \label{fig:KF_orig} 
\end{figure}

The common approach to Kalman filter design is to use the $C_L$ predicted by the mG-K model as a prediction step and then combine it with the pressure based $C_L$ estimate in an update step as shown schematically in Fig. \ref{fig:KF_orig}.  In this approach, the state of the system is just $C_L$, the time-update equation for this system is
\begin{equation}\label{eq:dynamic}
    {C_{L,\textrm{mG-K}}}(t_{k+1}) = a_k {{\hat C}_L}(t_k) + b_k
\end{equation}
where ${\hat C}_L$ denotes the posterior state estimate state estimated from the Kalman filter given the measurement. The measurement matrix is simply 
\begin{equation}\label{eq:KF_H}
H=
\begin{bmatrix} 
    1  \\
\end{bmatrix}.
\end{equation} 

\begin{figure}[h]
	\centering
    \includegraphics[width=0.5\textwidth]{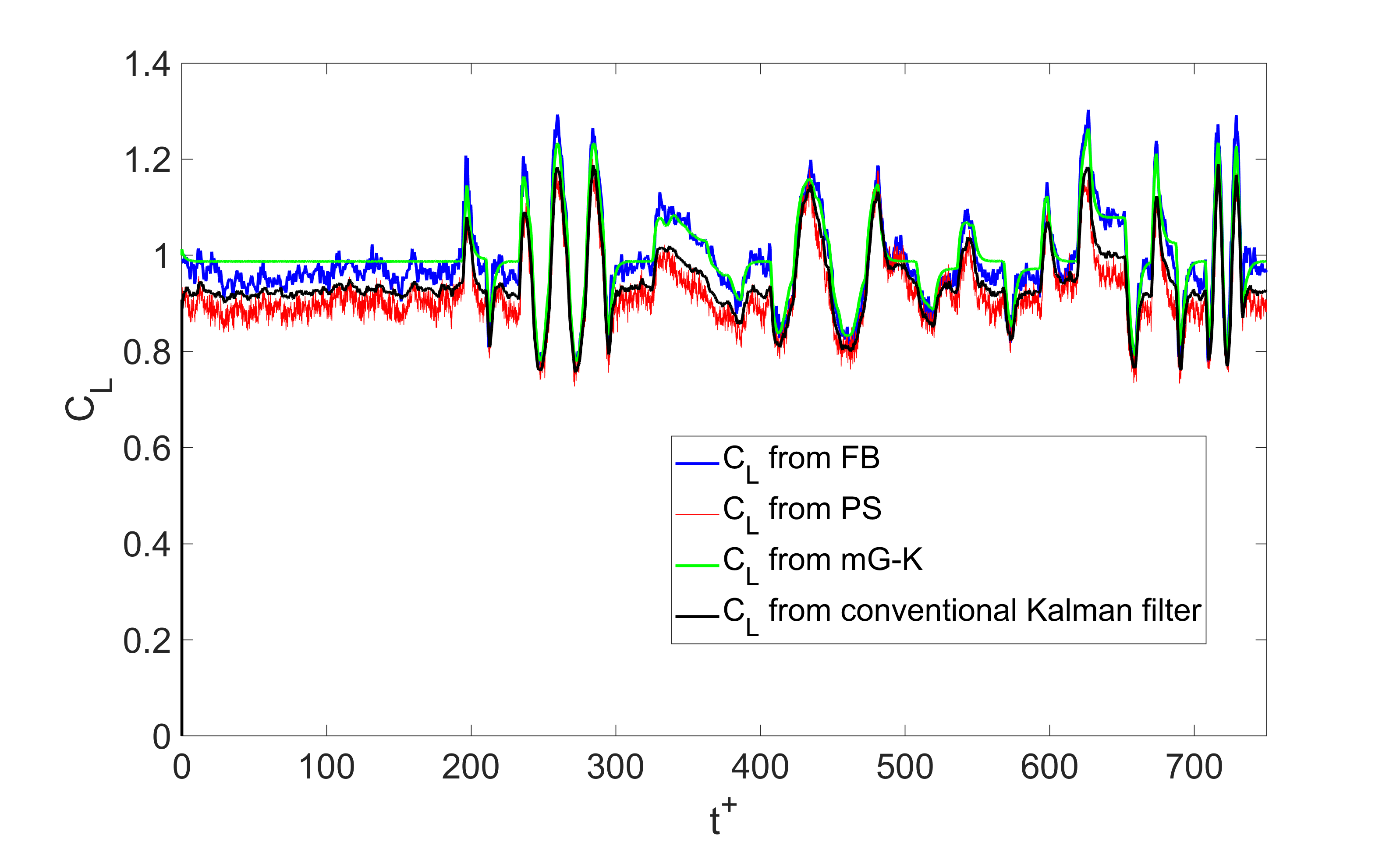}
    \caption{State estimation using the conventional Kalman filter for the first random pitching maneuver.} 
    
    \label{fig:KF_random_orig} 
\end{figure}

The time varying lift coefficient measured by the force transducer for the same quasi-random motion discussed in the last section is shown  in Fig. \ref{fig:KF_random_orig}, along with the $C_L$ predicted by the pressure measurements alone, the mG-K model alone, and the estimate from the combined Kalman filter.  

\begin{figure}[h]
	\centering
    \includegraphics[width=0.5\textwidth]{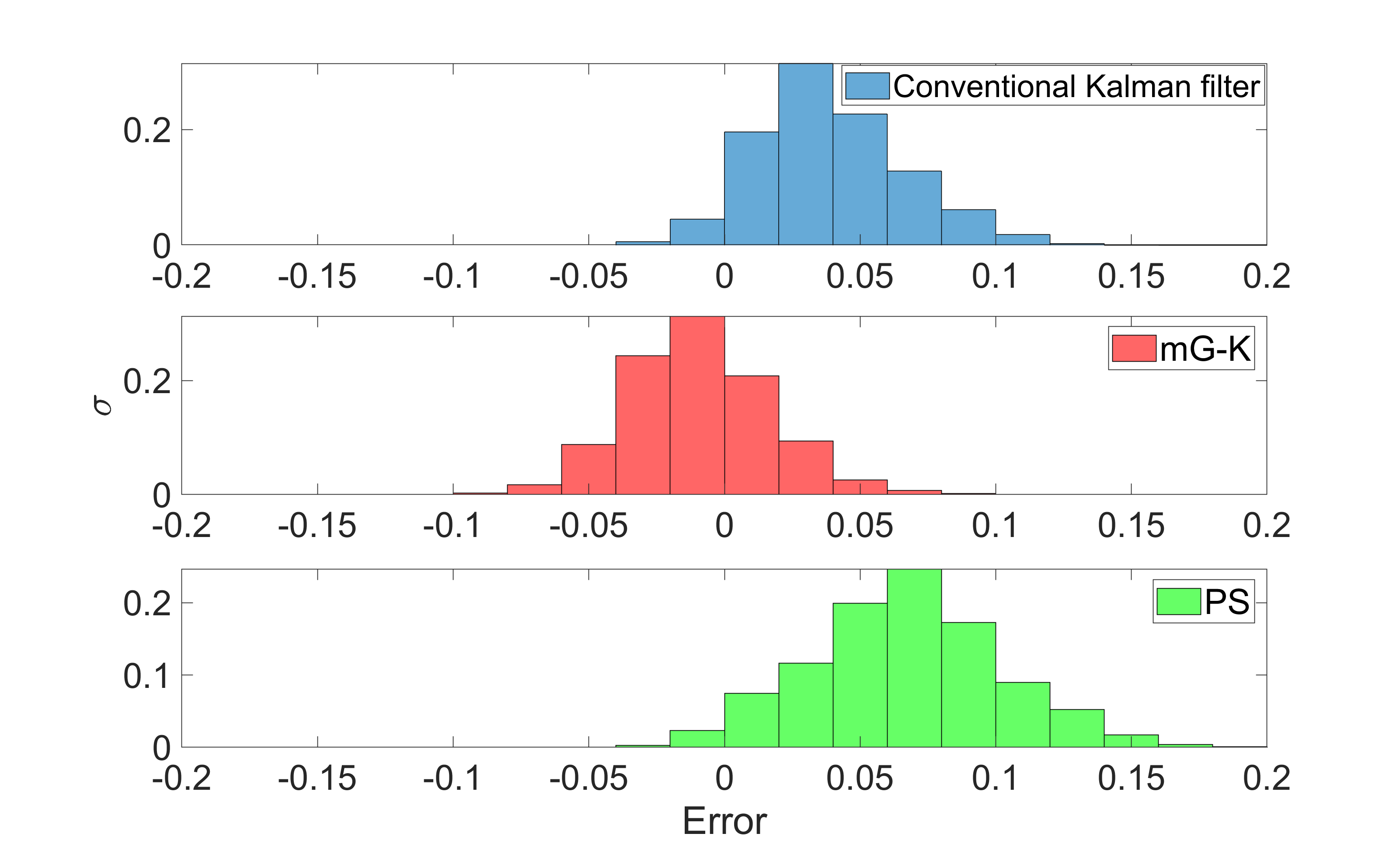}
    \caption{Error comparison of the conventional Kalman filter for the first random pitching case of Sec.~\ref{sec:LPV}.} 
    \label{fig:ERR_random_case1_classic} 
\end{figure}

The conventional Kalman filter design reduces the inherent noise in the pressure-based estimator, and while the filter's prediction bias (0.039) and rms error (0.048) are reduced from those of the pressure-based estimator alone (bias and rms error are 0.066 and 0.075, respectively). However, the bias of the conventional Kalman filter is worse than the mG-K model (-0.0094) by its own (see Fig.~\ref{fig:KF_random_orig}). The distribution of error in the Kalman filter estimated lift coefficient is compared to that solely from the pressure-inferred and mG-K models in Fig. \ref{fig:ERR_random_case1_classic}.

\FloatBarrier

\FloatBarrier

\FloatBarrier

We now discuss an improved Kalman filtering approach that utilizes the filter's ability to remove noise from the pressure measurement, while retaining its information about flow dynamics not captured by the mG-K model and allowing it to partially remove the bias associated with the pressure-inferred lift value.  To do this, we use the mG-K inferred lift value, together with the individual weights associated with each pressure sensor, to predict a model-based pressure for each sensor.  That is, we write 
\begin{equation}\label{eq:CL_pressure}
P'_j(t_k)=\frac{C_{L,\textrm{mG-K}}(t_k)}{w_j}-\sum_{i=1, i \neq j}^{5} \frac{w_i}{w_j} P_i(k),
\end{equation}
where $P'_j$ is the model-predicted pressure for the $j$th sensor, given the other sensor's readings.  Note that we also include the offset pressure $P_{5}$ in the sum and predict a model-consistent offset, $P'_{5}$.  Next, we expand the state space of the model to
\begin{equation}
\mathbf{x} =
\begin{bmatrix} 
    {C_L} &
    {P'_1} &        
    {P'_2} &
    {P'_3} &
    {P'_4} &
    {P'_5} &
\end{bmatrix}^T,
\end{equation}
and use the expanded predictor step
\begin{equation}
    \mathbf{x}(t_{k+1}) = \mathbf{a}_k \hat{\mathbf{x}}(t_k) + \mathbf{b}_k,
\end{equation}
where
\begin{equation}\label{eq:KF_p}
\mathbf{a}_k=
\begin{bmatrix} 
    a_k & 0 & 0 & 0 & 0 & 0        \\
    \frac{1}{w_1} & 0 & -\frac{w_2}{w_1} & -\frac{w_3}{w_1} & -\frac{w_4}{w_1} & -\frac{w_5}{w_1}        \\
    \frac{1}{w_2} & -\frac{w_1}{w_2} & 0 & -\frac{w_3}{w_2} & -\frac{w_4}{w_2} & -\frac{w_5}{w_2}        \\
    \frac{1}{w_3} & -\frac{w_1}{w_3} & -\frac{w_2}{w_3} & 0 & -\frac{w_4}{w_3} & -\frac{w_5}{w_3}        \\
    \frac{1}{w_4} & -\frac{w_1}{w_4} & -\frac{w_2}{w_4} & -\frac{w_3}{w_4} & 0 & -\frac{w_5}{w_4}        \\
    \frac{1}{w_5} & -\frac{w_1}{w_5} & -\frac{w_2}{w_5} & -\frac{w_3}{w_5} & -\frac{w_4}{w_5} & 0        \\
\end{bmatrix},
\end{equation}  
and
\begin{equation}
\mathbf{b}_k =
\begin{bmatrix} 
    b_k &
    0 &        
    0 &
    0 &
    0 &
    0 &
\end{bmatrix}^T.
\end{equation}
Finally, we write the measurement matrix
\begin{equation}\label{eq:KF_H}
H=
\begin{bmatrix} 
    0 & 1 & 0 & 0 & 0 & 0        \\
    0 & 0 & 1 & 0 & 0 & 0        \\
    0 & 0 & 0 & 1 & 0 & 0        \\
    0 & 0 & 0 & 0 & 1 & 0        \\
\end{bmatrix},
\end{equation} 
where there are only four outputs from the 6-dimensional state space corresponding to the four pressure sensors.

\begin{figure}[h]
	\centering
    \includegraphics[width=0.45\textwidth]{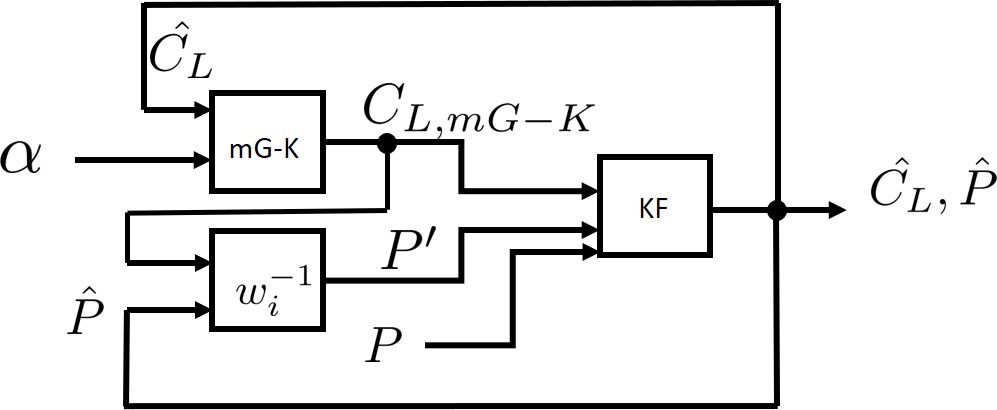}
    \caption{Schematic of the improved Kalman filter design.} 
    \label{fig:KF} 
\end{figure}

The architecture of the newly proposed (improved) Kalman filter is shown in Fig. \ref{fig:KF}.
We define $\boldsymbol\omega$ to be the process noise (for each of the 6 state variables) and $\boldsymbol\nu$ as the measurement noise (for each of the 4 sensors). The associated covariance matrices are $Q=E(\boldsymbol\omega \boldsymbol\omega^T)$ and $R=E(\boldsymbol\nu \boldsymbol\nu^T)$, which are chosen to be diagonal matrices with equal diagonal entries. The values of the diagonal entries of $R$ matrix could be $10^3$ to $10^4$ times larger than $Q$ to reduce the measurement noise. The standard Kalman filter procedure using a prediction step and measurement update is followed here. 

The main advantage of this improved Kalman filter algorithm  compared to the original one is that the pressure is directly coupled with the mG-K model through the shared value of $\hat{C_L}$.  This helps reduce the influence of the pressure-based biased error on the estimated $C_L$. 
\FloatBarrier 

\section{Validation of the improved Kalman filter} \label{sec:validate}

\begin{figure}[h]
	\centering
    \includegraphics[width=0.5\textwidth]{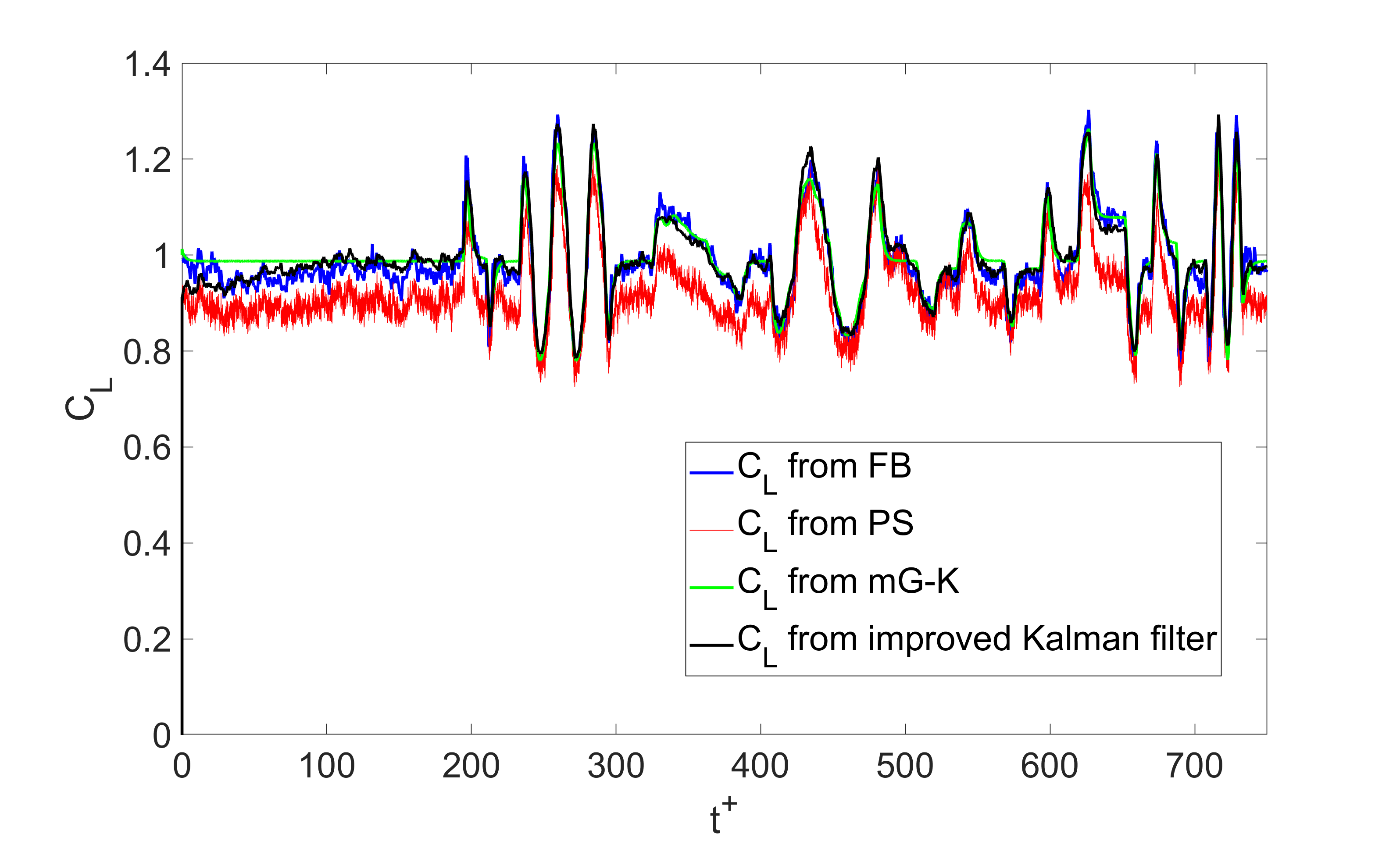}
    \caption{State estimation using the improved Kalman filter for the first random pitching maneuver.} 
    \label{fig:KF_random} 
\end{figure}

\begin{figure}[h]
	\centering
    \includegraphics[width=0.5\textwidth]{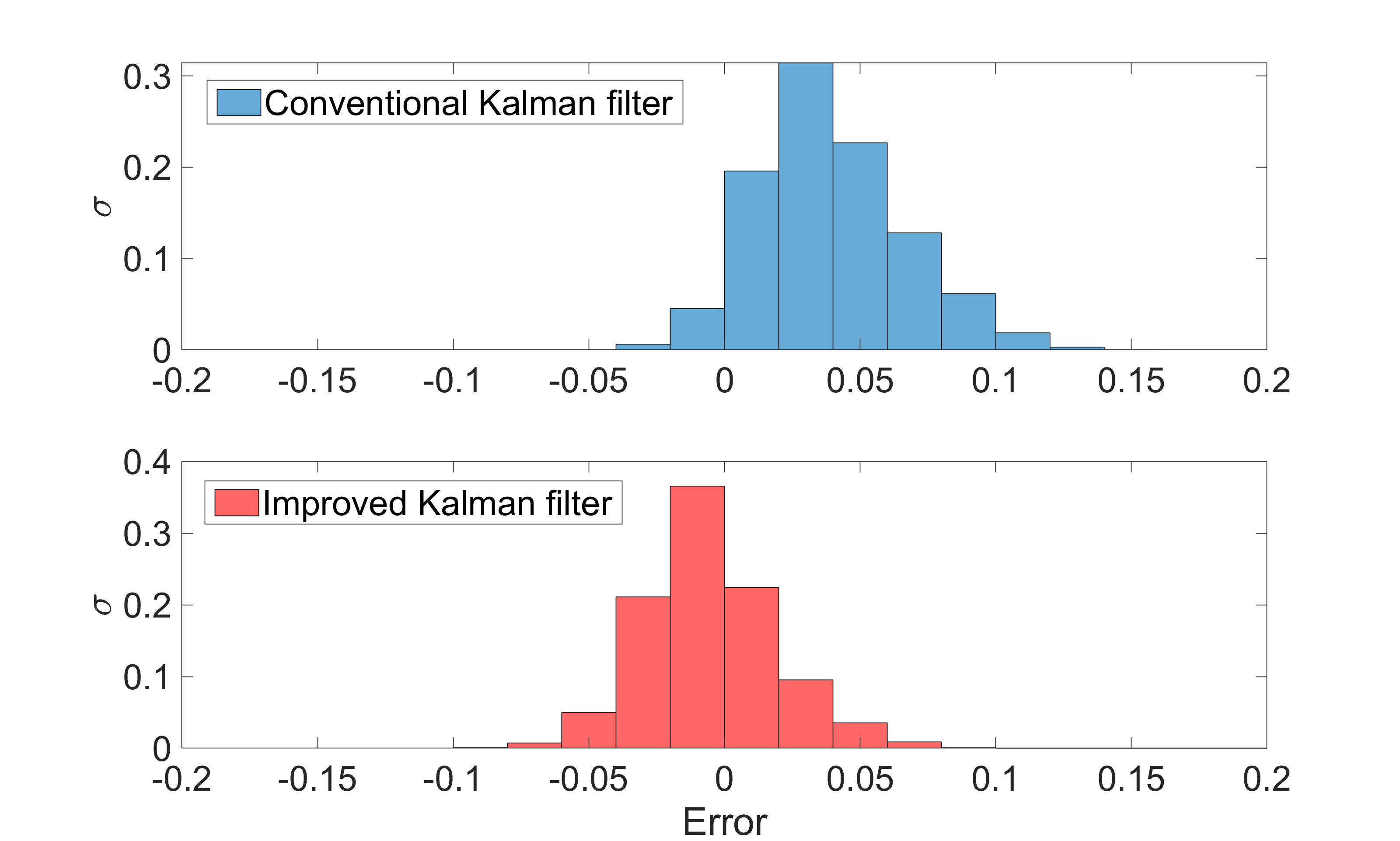}
    \caption{Error comparison between the conventional Kalman filter and the improved Kalman filter for the first random pitching case.} 
    \label{fig:KF_compare} 
\end{figure}

To validate the new filter, we first use the same random pitching motion discussed above.  The results are shown in Fig. \ref{fig:KF_random}. They show that the $C_L$ estimation now tracks the experimental force balance data for $C_L$ very well. The overall trend and even a portion of the detailed fluctuations in $C_L$ are captured, and the $C_L$ noise level is also reduced by the new  approach. The correlation coefficient between the experimental data and the improved Kalman filter output is 0.964, which is higher than either the mG-K model (0.956) or pressure $C_L$ estimation (0.928) alone. The mean bias is reduced from 0.039 to -0.0055 and the rms is reduced from 0.048 to 0.025 compared to the conventional Kalman filter. The error distribution comparison between the conventional and improved Kalman filter is shown in Fig. \ref{fig:KF_compare}. The performance improvement of the improved Kalman filter is so obvious.


\FloatBarrier 

\begin{figure}[h]
	\centering
    \includegraphics[width=0.5\textwidth]{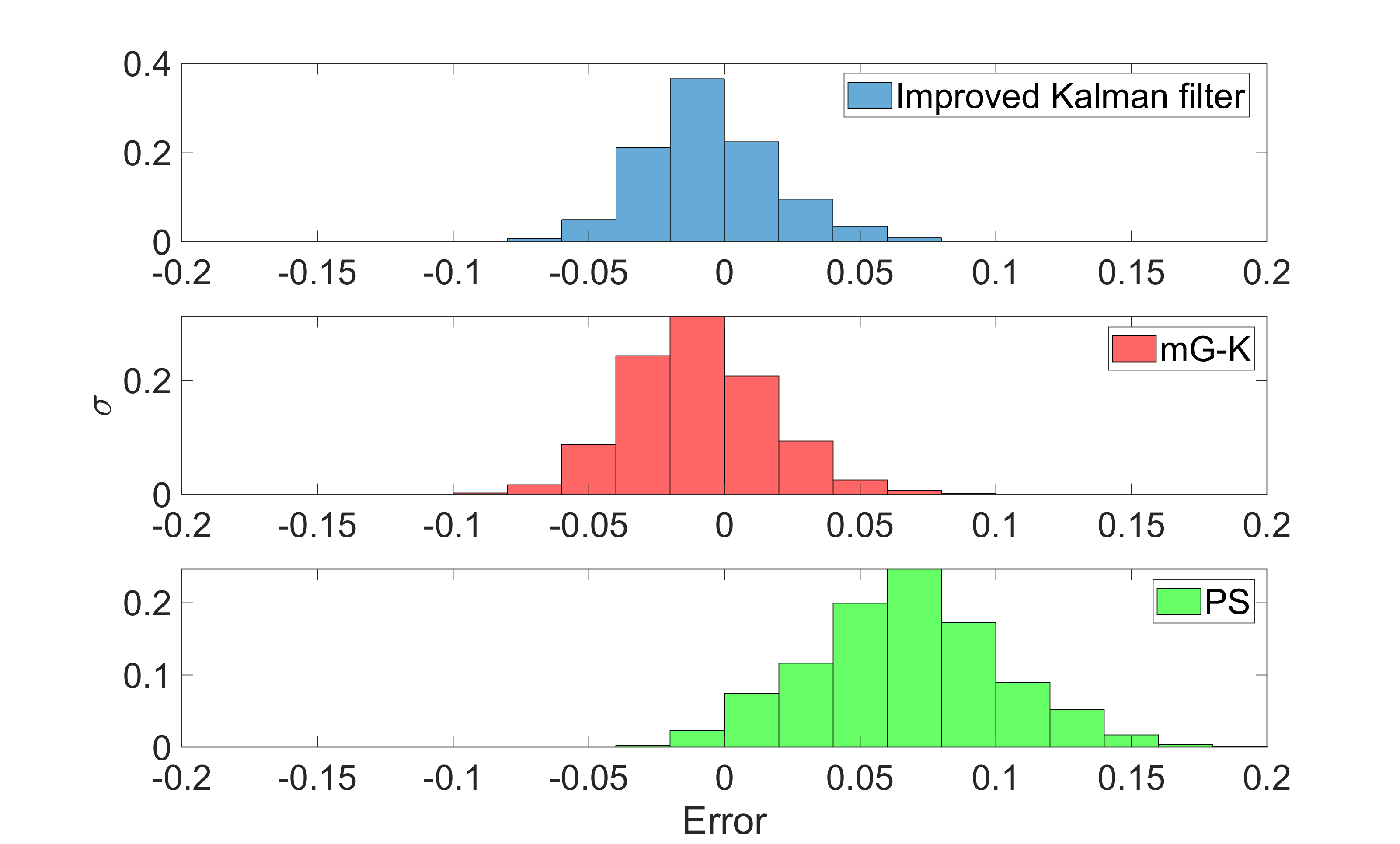}
    \caption{Error comparison of the improved Kalman filter for the first random pitching case of Sec.~\ref{sec:LPV}.} 
    \label{fig:ERR_random_case1} 
\end{figure}

To further illustrate the benefit of the new approach, the distribution of error in the lift coefficient is compared to that from solely from the pressure-inferred and mG-K models in Fig. \ref{fig:ERR_random_case1}. The improved Kalman filter significantly decreases the bias and reduces the variance in the estimates compared to the mG-K modeled and the weighted pressure on their own.

\FloatBarrier

\FloatBarrier

 
One could argue that the good performance of the improved Kalman filter is primarily due to the high accuracy of the mG-K model. To further test the ability of the improved Kalman filter, two types of artificial errors were added to the mG-K model. The first modeling error (case 1) is simulated through an error in which the $\dot{\alpha}$ input amplitude was reduced by 80\% from the actual value, and then multiplied by an additional error term of $\sin(t/0.05)$. The second modeling error (case 2) employs incorrect time constants in the mG-K model to simulate a time response error. In this case $\tau_1$ is increased by 55\% from its actual value and $\tau_2$ is reduced to 20\% of its actual value. 

\begin{figure}[h]
	\centering
    \includegraphics[width=0.5\textwidth]{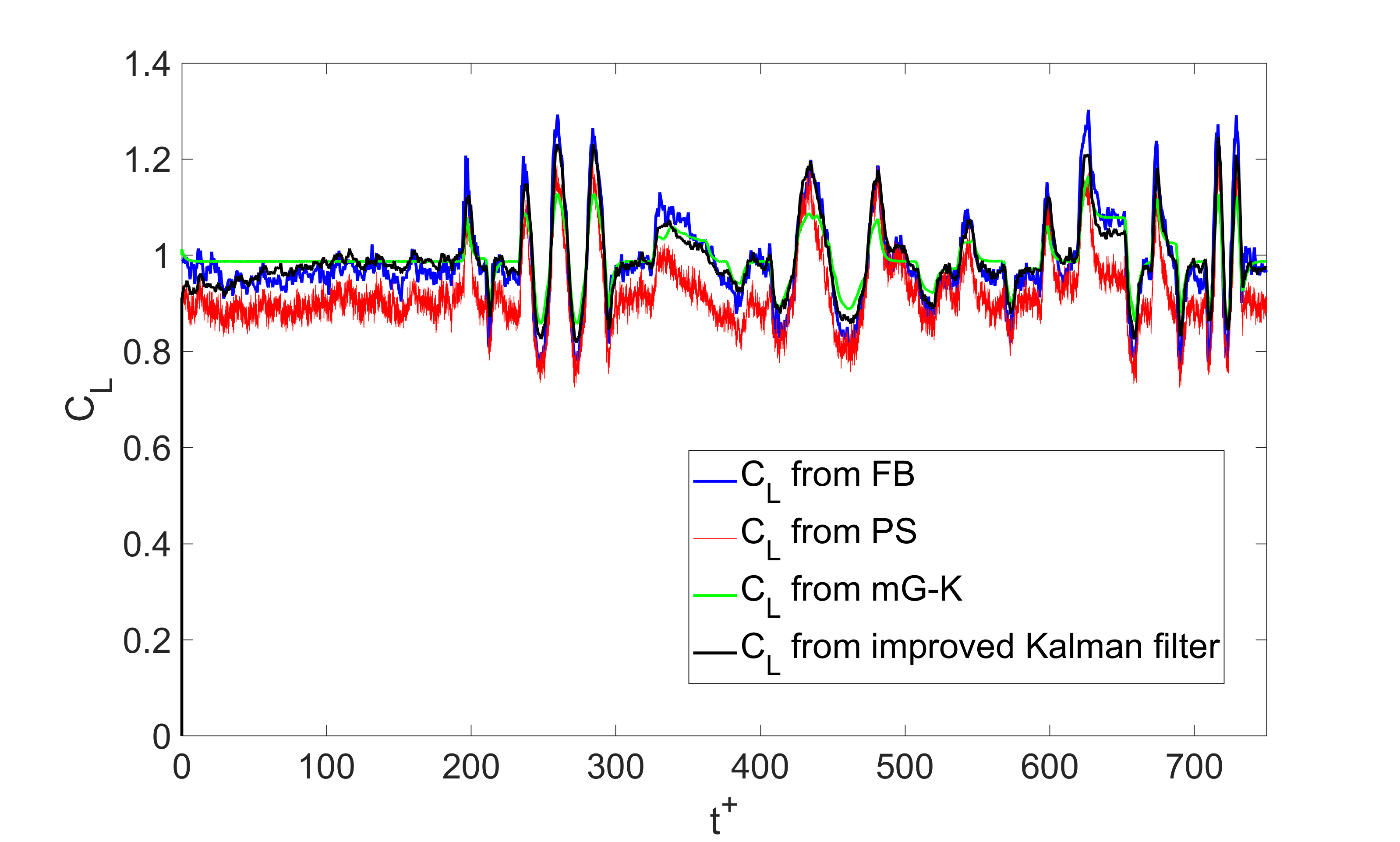}
    \caption{Improved Kalman filter with incorrect $\dot{\alpha}$ within the mG-K model (case 1) for the first random pitching maneuver.}
    \label{fig:KF_random_bad_GK_bad_amp} 
\end{figure}

\begin{figure}[h]
	\centering
    \includegraphics[width=0.5\textwidth]{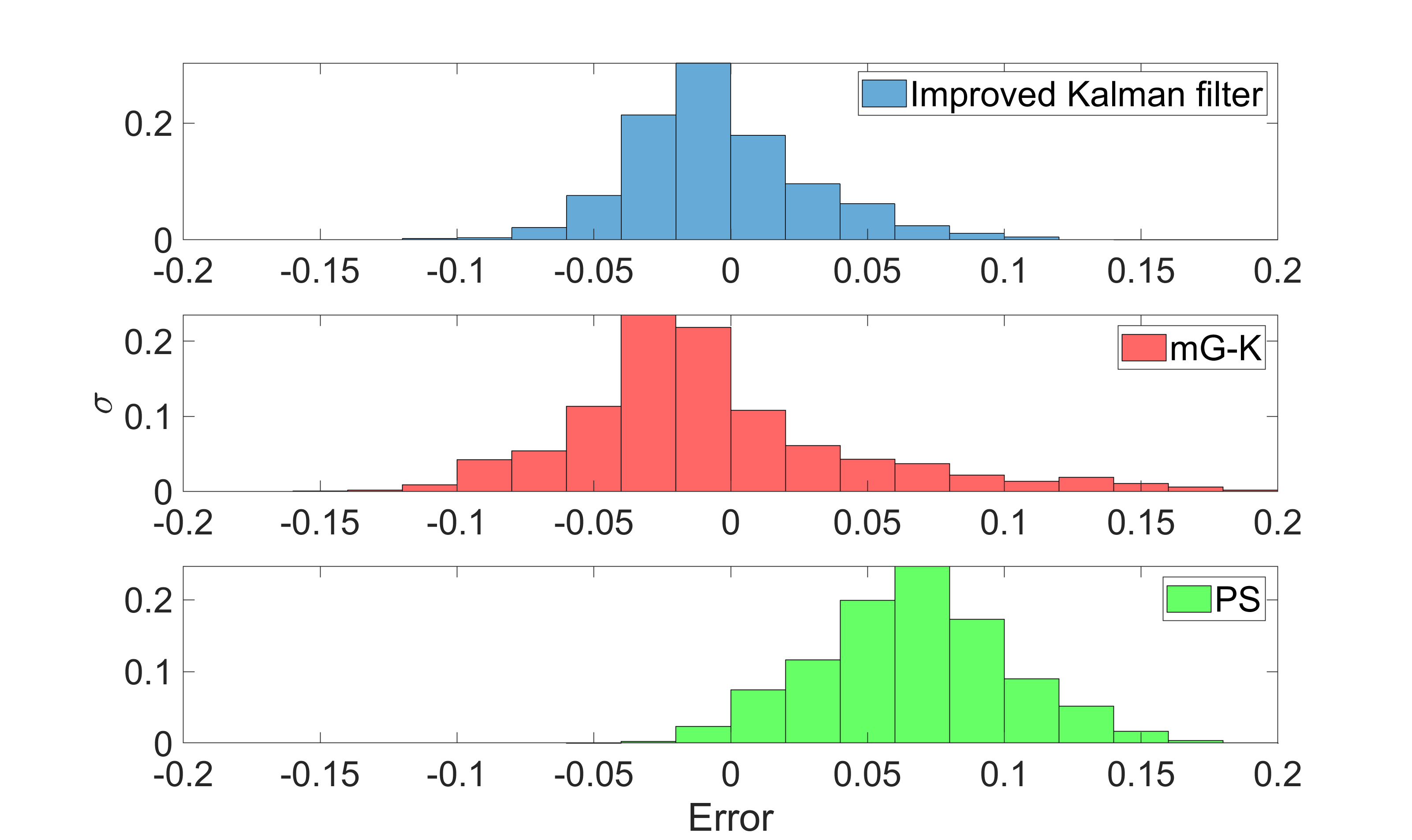}
    \caption{Error comparison of the improved Kalman filter for the first random pitching with wrong $\dot{\alpha}$ (case 1).} 
    \label{fig:ERR_random_case1_bad_alpha} 
\end{figure}

\begin{figure}[h]
	\centering
    \includegraphics[width=0.5\textwidth]{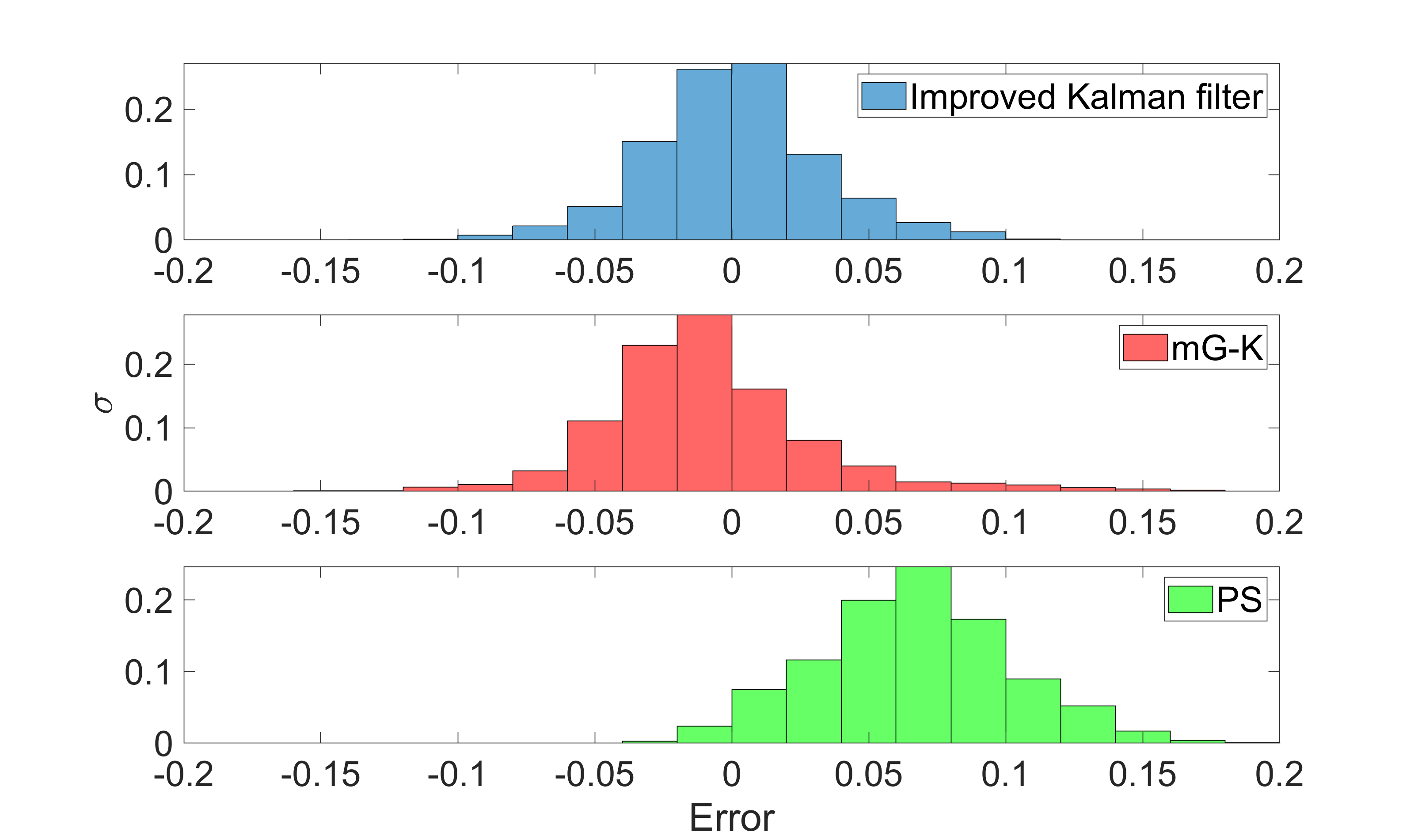}
    \caption{Error comparison of the improved Kalman filter for the first random pitching with wrong time constants (case 2).} 
    \label{fig:ERR_random_case1_bad_tau} 
\end{figure}

The results are shown in Fig. \ref{fig:KF_random_bad_GK_bad_amp} and Fig. \ref{fig:ERR_random_case1_bad_alpha} for case 1, and Fig. \ref{fig:ERR_random_case1_bad_tau} Fig. \ref{fig:ERR_random_case1_bad_tau} for case 2, respectively. As expected, in case 1 the rms error in the lift coefficient prediction by the mG-K model increases from 0.028 to 0.053 compared with the one with the right $\dot{\alpha}$. Even with this mG-K model error, the improved Kalman filter is still able to partially compensate for case 1 model error, and the lift coefficient estimation rms error increases by a smaller amount from 0.025 to 0.033 compared to the one with the right mG-K model. The correlation coefficients between the improved Kalman filter output and the experimental force balance data are reduced slightly from 0.964 to 0.949 for case 1 compared to the one with the right mG-K model.



For the time-constant error in case 2 the rms error in the mG-K model increases from  0.028 to  0.040 compared to the one with right time constants.  Again the improved Kalman filter is able to partially compensate for the modeling error, although the kalman filter rms error is increased from  0.025 to 0.032 compared to the one with the corrected mG-K model. The correlation coefficient between the measured $C_L$ and the improved Kalman filter estimate is 0.939 for case 2. 

In both cases, the improved Kalman filter tracks the force balance measured $C_L$ signal well. Both of the error histograms indicate that the improved Kalman filter is capable of reducing error associated with the model errors in the mG-K model. The improved Kalman filter outperforms the mG-K model and weighted pressure by their own in both cases. 

\FloatBarrier 

\FloatBarrier

\begin{figure}[h]
	\centering
    \includegraphics[width=0.5\textwidth]{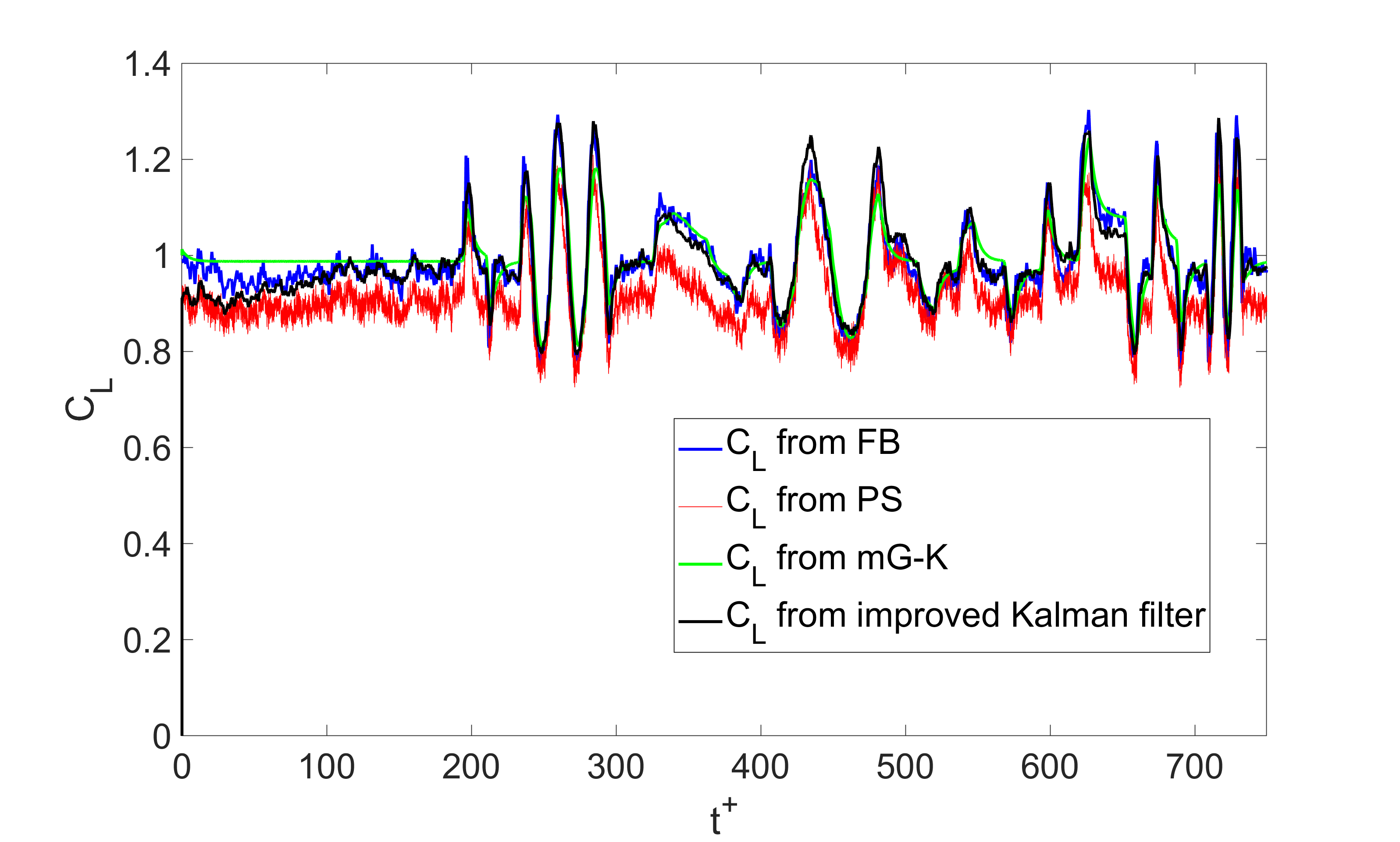}
    \caption{Improved Kalman filter with incorrect time constants within the mG-K model (case 2) for the first random pitching maneuver.} 
    \label{fig:KF_random_bad_GK_bad_tau} 
\end{figure}
\FloatBarrier 

\FloatBarrier

For additional validation of the improved Kalman filter, we tested it with the second random motion. The major difference between the second random pitching motion and the first one is the smaller pitching amplitude. A smaller pitching amplitude results in a lower signal-to-noise ratio for the mG-K model, because the effects of background turbulence and wake turbulence, which are not modeled by the mG-K model, play larger roles in the $C_L$ variation.

\begin{figure}[h]
	\centering
    \includegraphics[width=0.5\textwidth]{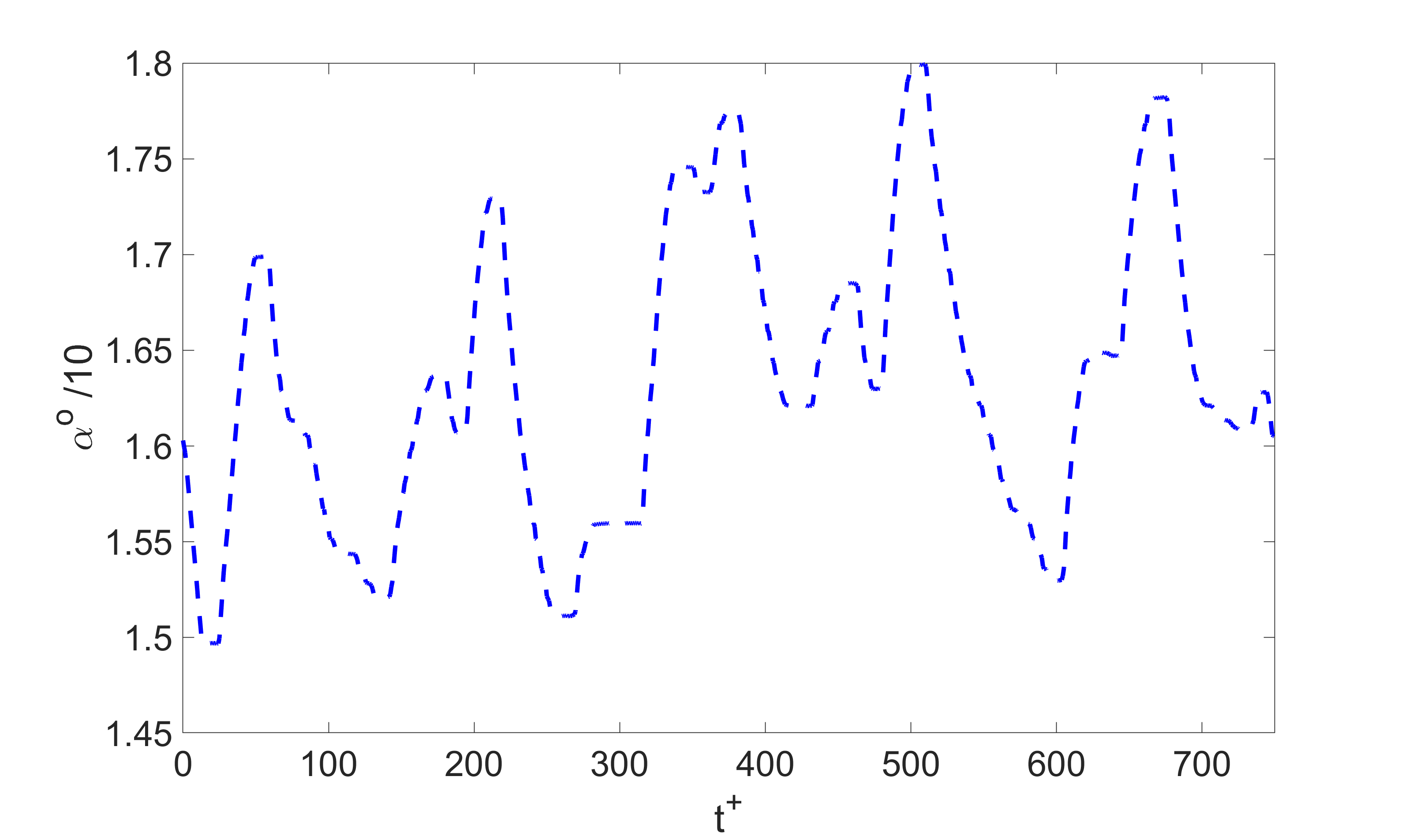}
    \caption{The time series of the $\alpha$ for the second quasi-random data set.} 
    \label{fig:alpha_random_case2} 
\end{figure}

\begin{figure}[h]
	\centering
    \includegraphics[width=0.5\textwidth]{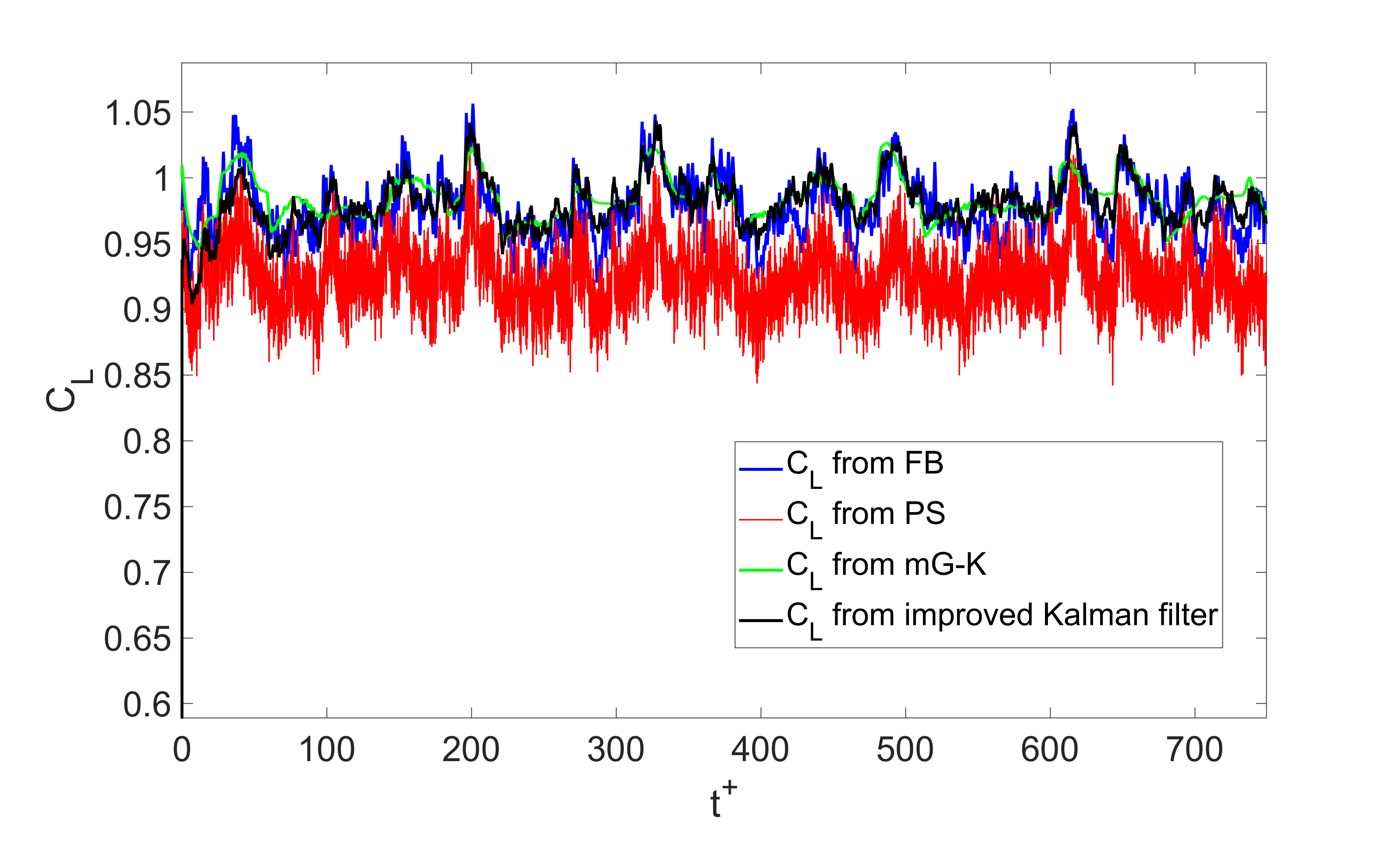}
    \caption{State estimation using the improved Kalman filter for the second random pitching maneuver.} 
    \label{fig:KF_random_case2} 
\end{figure}

Fig. \ref{fig:alpha_random_case2} shows the pitching motion of the second random pitching signal. The lift coefficient prediction of the improved Kalman filter using the second random pitching maneuver is shown in Fig. \ref{fig:KF_random_case2}. It can be seen that the improved Kalman filter outperforms both the mG-K model prediction and the weighted pressure estimates of $C_L$ by their own. To quantify the performance of the improved Kalman filter against the second random pitching maneuver, the correlation coefficient between the force balance measured $C_L$ and the Kalman filter is 0.707, which is higher than either the mG-K model (0.636) or pressure $C_L$ estimation (0.646) by its own. 

Power spectral density function comparisons of the force balance measured lift with the spectra output from two Kalman filtering approaches indicated that the Kalman filter generated signals had less energy at frequencies above 2 Hz in the first type of maneuver and above 0.5 Hz in the second type of maneuver. We suspect the low-order of the mG-K model is connected to the Kalman filter's inability to track high-frequency disturbances.  

\begin{figure}[h]
	\centering
    \includegraphics[width=0.5\textwidth]{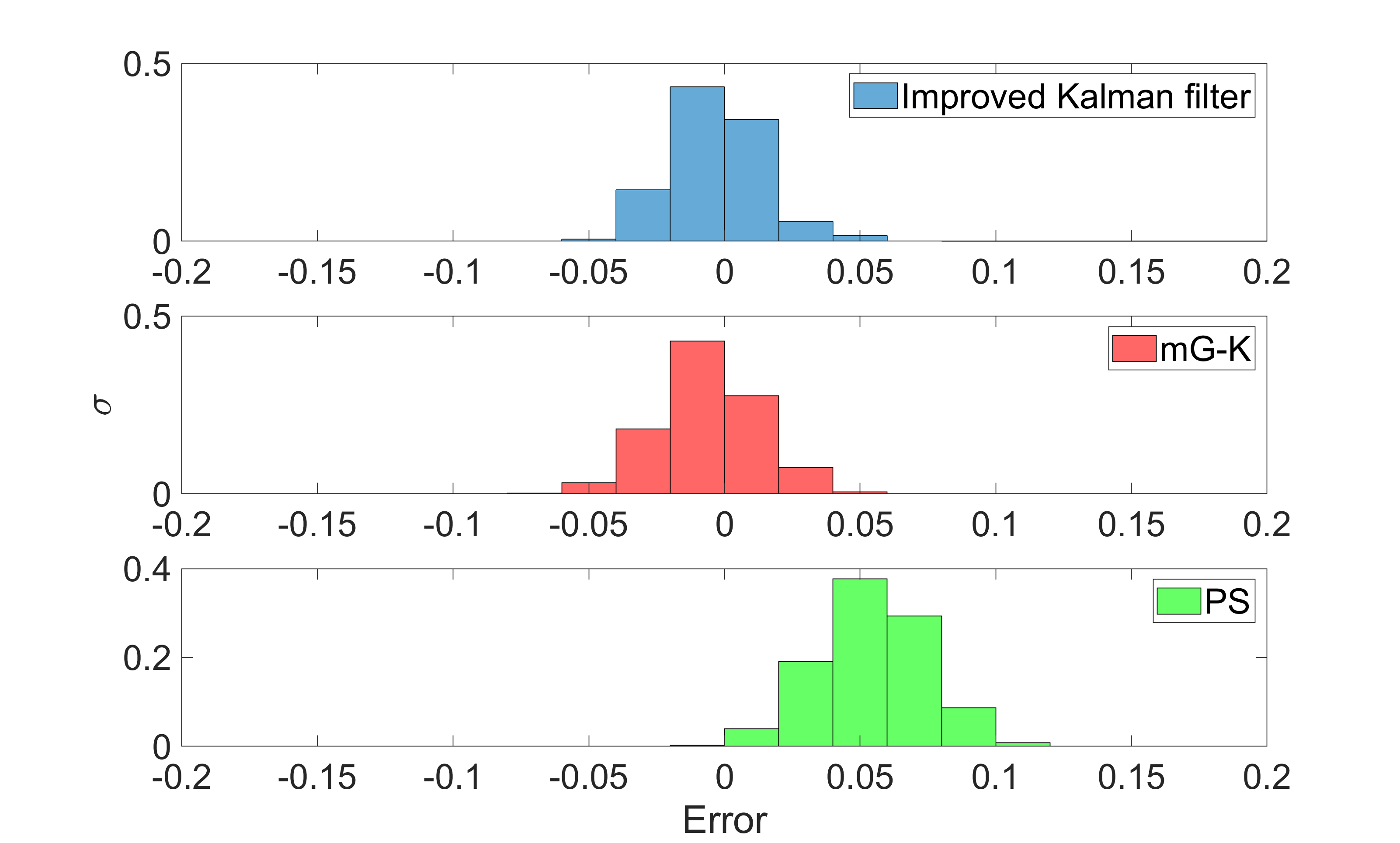}
    \caption{Error comparison of the improved Kalman filter for the second random pitching maneuver.} 
    \label{fig:ERR_random_case2} 
\end{figure}

The histogram of the error distribution for the second random pitching case is shown in Fig. \ref{fig:ERR_random_case2}. It further indicates that the improved Kalman filter is still capable of suppressing both the white noise (produced by the mG-K model and the pressure measurements) and colored noise (produced by the pressure measurements) even if the $C_L$ cannot be modeled accurately by the mG-K model. The biased error is largest with the pressure based estimate (0.054), and the Kalman filter reduces it to -0.003.  The rms error in the pressure-based estimate is 0.058, and again the Kalman filter reduces that rms error to 0.017.  



\FloatBarrier

\FloatBarrier

\FloatBarrier

\section{Conclusion}\label{Sec:conc} A method for improving the accuracy of instantaneous lift coefficient estimates on an NACA-0009 airfoil undergoing random pitching maneuvers is demonstrated.  The method introduces a novel Kalman filter algorithm that assimilates a small number of surface pressure measurements with a modified Goman-Khrabrov model \textcolor{black}{that requires an angle of attack measurement}. The $C_L$ estimation based on four surface pressure measurements \textcolor{black}{and angle of attack} was obtained using a weighted average of the four pressures and an offset. Without using the Kalman filter the error histograms of the pressure-based $C_L$ estimate are shown to be biased. To compensate for the bias and to reduce the standard deviation of the error, a modified Goman-Khrabrov model was assimilated with the pressure data using a conventional Kalman filtering scheme. A necessary intermediate step required us to show that the modified Goman-Khrabrov model is in fact a linear parameter varying system with a nonlinear input forcing term, and hence it may be used in the Kalman filter.

Only a partial reduction in the bias and rms errors was achieved with the conventional Kalman filter approach, so the Kalman filter algorithm was modified to provide filtered estimates of both the lift coefficient and the input pressure signals. Better estimates of $C_L$ estimation were obtained with the improved Kalman filter approach.  The modeling error, the measurement error, the bias, and the noise were all reduced by the final modification. By including additional modeling errors in pitch rate and mG-K model time constants, we show that the improved Kalman filtering approach still provides accurate estimates of the time-varying $C_L$.

\section*{Appendix}
\section*{Proof that the Kalman filter is applicable to LPV systems with nonlinear input}\label{Sec:A}

An LPV dynamic system with nonlinear input can be written as
\begin{equation} \label{eq:nlpv}
X_{k+1}=A_kX_{k}+F(u_{k})+\omega_{k}
\end{equation}
where $X_k$ is the state $X\in \mathbb{R}^n$ at time instant $k$, $F(u)$ is the nonlinear input function and $\omega$ is the white Gaussian processing noise. The measurement $Z\in\mathbb{R}^m$ at time instant $k+1$ is
\begin{equation} \label{eq:measrue}
Z_{k+1}=HX_{k+1}+\nu_{k+1}
\end{equation}
where $H$ is the measurement matrix and $\nu$ is the measurement noise. Following a similar algorithm proposed by Kalman, et. al. \cite{kalman1960new} and Welch and Bishop \cite{welch1995introduction}, the time update of the discrete Kalman filter can be then expressed as 
\begin{equation} \label{eq:state_update}
\hat{X}^-_{k+1}=A_k\hat{X}_k+F(u_{k})
\end{equation}
\begin{equation} \label{eq:state_p}
P^-_{k+1}=E[e^-_{k+1}e^{-T}_{k+1}].
\end{equation}
Here, $\hat{X}^-_{k+1}$ is the \textit{a priori} estimate at $(k+1)th$ time step, $P^-_{k+1}$ is the \textit{a priori} estimate error covariance, $e_{k+1}^-$ is the \textit{a priori} estimate error, $E[.]$ denotes the expected value.  $\hat{X}_k$ is the \textit{a posteriori} state estimate at time step $k$ as a linear combination of the \textit{a priori} estimate $\hat{X}^-_k$ and a weighted difference between the actual measurement $Z_k$ and a measurement prediction $H\hat{X}^-_k$. Hence,  the discrete Kalman filter measurement update equations are
\begin{equation} \label{eq:K}
K_{k+1}=P^-_{k+1}H^T(HP_{k+1}^-HT+R)^{-1}
\end{equation}
\begin{equation} \label{eq:X}
\hat{X}_{k+1}=\hat{X}^-_{k+1}+K_{k+1}(Z_{k+1}-H\hat{X}^-_{k+1})
\end{equation}
\begin{equation} \label{eq:P}
P_{k+1}=(I-K_{k+1}H){P}^-_{k+1}
\end{equation}
where $K_{k+1}$ is the Kalman gain at time instant $k+1$, $R$ is the measurement noise covariance and $P_{k+1}$ is the \textit{a posteriori} estimate error covariance. 

At this stage, $P_{k+1}^-$ in Eq.~(\ref{eq:state_p}) is the only term that contains the nonlinear input function $F(u_k)$.  The \textit{a priori} estimate error $e^-_{k+1}$ can be expressed as
\begin{align} \label{eq:e}
e^-_{k+1} & =X_{k+1}-\hat{X}^-_{k+1} \\
& =[A_{k}X_{k}+F(u_{k})+\omega_{k}]-[A_k\hat{X}^+_k+F(u_{k})],
\end{align}
and it can be seen that the $F(u_k)$ terms are canceled out. Thus, the \textit{a priori} estimate error covariance becomes
\begin{align} \label{eq:pp}
P^-_{k+1} & =E[e^-_{k+1}e^{-T}_{k+1}] \\
& =A_kP_kA_k^T+Q 
\end{align}
where $Q$ is process noise covariance. By replacing Eq.~ (\ref{eq:state_p}) with Eq.~(\ref{eq:pp}), it can be seen that the Kalman filter algorithm for the LPV dynamic system with nonlinear input is the same as the original Kalman filter despite the time-varying $A_k$. 

\begin{acknowledgements}
The support for this work by the US Air Force
Office of Scientific Research FA9550-18-1-0440 with program
manager Gregg Abate is gratefully acknowledged. 
Supported under the US Air Force
Office of Scientific Research FA9550-14-1-0328 with program manager Douglas Smith is also gratefully acknowledged. The first author would like to thank the supported under ONR MURI Grant N00014-14-1-0533 with program manager Robert Brizzolara. The insightful suggestions from Professor C. Rowley and S. Otto at Princeton University are sincerely appreciated.

\end{acknowledgements}

%
\section*{Conflict of interest}

The authors declare that they have no conflict of interest.

\bibliographystyle{spmpsci}      

\bibliography{mybib}


\end{document}